\documentclass[aps, reprint, superscriptaddress, prx, preprintnumbers, amsmath, amssymb, letterpaper, floatfix, showpacs]{revtex4-2}
\usepackage{graphicx, color}
\begin{document} 

\title{Intrinsic new properties of a quantum spin liquid}

\author{Y. X. Yang}
\affiliation{State Key Laboratory of Surface Physics, Department of Physics, Fudan University, Shanghai 200433, China}
\author{Xin Li}
\affiliation{State Key Laboratory of Surface Physics, Department of Physics, Fudan University, Shanghai 200433, China}
\affiliation{Key Laboratory of Neutron Physics and Institute of Nuclear Physics and Chemistry, China Academy of Engineering Physics (CAEP), Mianyang 621999, China} 
\author{C. Tan}
\author{Z. H. Zhu}
\author{J. Zhang}
\author{Z. F. Ding}
\author{Q. Wu}
\author{C. S. Chen}
\affiliation{State Key Laboratory of Surface Physics, Department of Physics, Fudan University, Shanghai 200433, China}
\author{T. Shiroka}
\affiliation{Laboratory for Muon-Spin Spectroscopy, Paul Scherrer Institut, 5232 Villigen, Switzerland}
\author{Y. H. Xia}
\affiliation{Key Laboratory of Neutron Physics and Institute of Nuclear Physics and Chemistry, China Academy of Engineering Physics (CAEP), Mianyang 621999, China} 
\author{D. E. MacLaughlin}
\affiliation{Department of Physics and Astronomy, University of California, Riverside, California 92521, USA}
\author{C. M. Varma}
\email[Corresponding author:]{chandra.varma@ucr.edu}
\affiliation{Department of Physics, University of California, Berkeley, California 94704, USA}
\author{L. Shu}
\email[Corresponding author:]{leishu@fudan.edu}
\affiliation{State Key Laboratory of Surface Physics, Department of Physics, Fudan University, Shanghai 200433, China}
\affiliation{Collaborative Innovation Center of Advanced Microstructures, Nanjing 210093, China}
\affiliation{Shanghai Research Center for Quantum Sciences, Shanghai 201315, China}

\date{\today}

\preprint{ver.7.2}

\begin{abstract} 
Quantum fluctuations are expected to lead to highly entangled spin-liquid states in certain two-dimensional spin-1/2 compounds. We have synthesized and measured thermodynamic properties and muon spin relaxation rates in the copper-based two-dimensional triangular-lattice spin liquids Lu$_3$Cu$_2$Sb$_3$O$_{14}$ and Lu$_3$CuZnSb$_3$O$_{14}$. The former is the least disordered of this kind discovered to date. Magnetic entropy generation at high temperatures has been ruled out after carefully correcting for the lattice specific heat. Surprisingly, roughly half of the magnetic entropy is missing down to temperatures of \textit{O}(10$^{-3}$) the exchange energy, independent of magnetic field up to $g\mu_B H \gtrsim k_B\Theta_W$, where $\Theta_W$ is the Weiss temperature. The magnetic specific heat divided by temperature $C_M(T)/T$ and muon spin relaxation rate $\lambda(T)$ are both temperature-independent at low temperatures, followed by logarithmic decreases with increasing temperature. This behavior can be simply characterized by scale-invariant time-dependent fluctuations with a single parameter. Since no cooperative effects due to impurities are observed, the measured properties are intrinsic. They are evidence that in Lu$_3$Cu$_2$Sb$_3$O$_{14}$ massive quantum fluctuations lead to either a gigantic specific heat peak from singlet excitations at very low temperatures or, perhaps less likely, an extensively degenerate possibly topological singlet ground state. 
\end{abstract}

\maketitle
~
\vspace{30pt} \section{INTRODUCTION} \label{sec:intro}

The study of quantum fluctuations in interacting matter is of primary interest in many fields of physics, encompassing fields as diverse as the thermodynamics of black holes \cite{Carlip2015, Maldacena18u}, particle physics beyond the standard model \cite{Ellis2009}, the theory of quantum computation \cite{Li11u}, and a considerable number of phenomena in condensed matter physics. The latter allow access and control to a wide variety of experiments, and the concepts often cut across different fields. These range from quantum Hall effects \cite{PrGi90} to the quantum criticality that governs high temperature superconductivity \cite{Varma_IOPrev2016, VarmaRMP2020} to spin-liquid states \cite{BCKN20, Savary16}, all of which have been intensively studied in the last three decades. 

Spin liquids, in particular, have been hard to characterize beyond the fact that quantum fluctuations prevent any conventional order in them. Despite extensive experiments, few precise conclusions about the nature of the ground state and low-lying excitations are available, because the results are almost always dominated by cooperative effects, however interesting, of the impurities \cite{Savary16, KTMK18, BCKN20}. 

We have synthesized the $S{=}1/2$ trigonal-lattice compounds Lu$_3$Cu$_2$Sb$_3$O$_{14}$ (LCSO) and Lu$_3$CuZnSb$_3$O$_{14}$ (LCZSO). They are variations on the $R_3$Zn$_2$Sb$_3$O$_{14}$ ($R$ = rare earth) series of tripod-kagom\'e-lattice compounds~\cite{LHWK14, SKC16, POHM16, DTLC17, DBPH20}, with $R$ = Lu and Zn completely (LCSO) or half (LCZSO) substituted by Cu. In LCSO the Cu ions form 2D triangular sublattices in two separate layers. It is remarkably defect-free: we estimate the magnetic impurity concentration to be $\lesssim$0.1\% and other impurity or defect concentrations $\sim$1\%. In LCZSO Cu and Zn ions predominantly occupy alternate triangular layers, but $\sim$5\% Cu/Zn site interchanges affect its properties compared to LCSO. 

Thermodynamic properties and muon spin relaxation ($\mu$SR) rates of these compounds have been measured, the latter down to 16~mK\@. There are no signatures of static magnetism, ordered or disordered, or any other cooperative effects of impurities in either compound, from 300~K down to the lowest temperatures. A very surprising result is that in LCSO the \emph{measured} magnetic entropy~$S_M(T)$ above 0.1~K, obtained from the specific heat and magnetization (the latter from Maxwell relations), saturates at high temperatures at roughly 40\% of the total magnetic entropy~$k_B\ln{2}$ per spin-1/2; $\sim$60\% of the magnetic entropy is \emph{missing}. $\mu$SR measurements effectively extend these results down to 16~mK\@. The missing entropy is independent of magnetic field up to 9~T\@. 

Furthermore, in both LCSO and LCZSO the magnetic specific heat coefficient $C_M/T$ and $\lambda(T)$ track each other; both are constant at low temperatures, followed by logarithmic decreases with increasing temperature. These results are shown to be consistent with scale-invariant magnetic fluctuations. 

The central result of our experiments is that the missing entropy resides at very low temperatures, or perhaps in the ground state. This requires accurate determination of the temperature dependence of $S_M(T)$ over a very wide range of temperatures. Up to about 20~K, which is of the order of the paramagnetic Weiss temperature~$\Theta_W$ from the susceptibility, the measured total specific heat of magnetic LCSO is clearly larger than that of nonmagnetic LZSO\@. Then $S_M(T)$ is obtained by subtraction, the accuracy of which is tested by measurement of the change in entropy in a magnetic field. This procedure yields $S_M(H{=}0,T)$ from 0.1~K to 20~K of only $0.36\,k_B\ln{2}$ per spin. 

At higher temperatures determination of $S_M(T)$ by subtraction becomes difficult, due to the small difference between the lattice specific heats~$C_\mathrm{latt}$ in the two compounds: their masses and force constants are not exactly the same. This difference is accounted for by scaling the temperature dependence of $C_\mathrm{latt}$ in LZSO before subtraction. The temperature scaling index~$\eta$ is a free parameter at this stage, subject to an entropy difference~$S_\mathrm{diff}(T)$ that must exhibit zero or positive slope and remain $\leq k_B\ln{2}$ per spin at high temperatures if it is indeed $S_M(T)$.

Complementary information on the magnetic entropy at higher temperatures has been obtained using reliable high-temperature expansions for the entropy and susceptibility of a Heisenberg model on the triangular lattice with nearest-neighbor and 2nd-nearest-neighbor exchange interactions~$J_1$ and $J_2$, respectively~\cite{Sing22}. The expansions are expected to yield reliable results from asymptotically high temperatures down to roughly $2J_1$ if $J_2 \ll J_1$. An accurate fit of the susceptibility expansion to the data determines $J_1$ and $J_2$, which are then used to deduce the entropy from its expansion. The best estimate of $\eta$ is obtained from the requirement that the scaled entropy at high temperatures agrees with the series expansion to within a constant. 

For $H = 0$ this procedure yields a constant magnetic entropy ${\sim}0.4k_B\ln{2}$ per spin above $\sim$25~K\@. Additional entropy generation at high temperatures is ruled out, and the missing ${\sim}0.6k_B\ln{2}$ per spin must therefore arise from states below $\sim$0.1~K that do not contribute to the specific heat above that temperature. This``ground-state'' entropy (more likely from very low-lying excited states) is a constant contribution at all experimental temperatures.

We believe it is the high purity of LCSO that allows us to characterize its extraordinary intrinsic properties, and that these properties are representative of a new class of spin liquids to which it belongs.

The paper is organized as follows. Structure determination and results of specific heat, entropy, magnetization, and $\mu$SR experiments are reported in Sec.~\ref{sec:results}, with details given in the Supplemental Information (SI)\@. In Sec.~\ref{sec:disc} we discuss high-temperature series expansions for thermal quantities, Cu-ion coordination and symmetry, evidence for two separate sublattices that are at most weakly coupled magnetically, a scale-invariant phenomenology, theories of spin liquids, and comparison of our results with other spin liquids. Section \ref{sec:concl} summarizes the evidence for colossal quantum fluctuations in LCSO, which result in completely new ground and low-lying excited states of a spin liquid.

\section{RESULTS} \label{sec:results}

\subsection{Structure} \label{sec:struct}

LCSO, LCZSO, and the isostructural nonmagnetic compound Lu$_3$Zn$_2$Sb$_3$O$_{14}$ (LZSO) belong to the tripod-kagom\'e-lattice family~\cite{LHWK14, SKC16, POHM16, DTLC17, DBPH20}, in which kagom\'e lattices are formed by alternating layers of filled-shell ($S{=}0$) Sb$^{5+}$ and rare-earth/Lu$^{3+}$ ions. The compounds were synthesized by solid-state reaction. Stoichiometric amounts of Lu$_2$O$_3$, CuO (ZnO for LCZSO and LZSO), and Sb$_2$O$_3$ were thoroughly mixed using an agate mortar, and heated to 1030$^{\circ}$C for 60 hours with intermediate regrinding and reheating. So far only powder samples could be prepared. 

\begin{figure*}
\includegraphics[width=\textwidth]{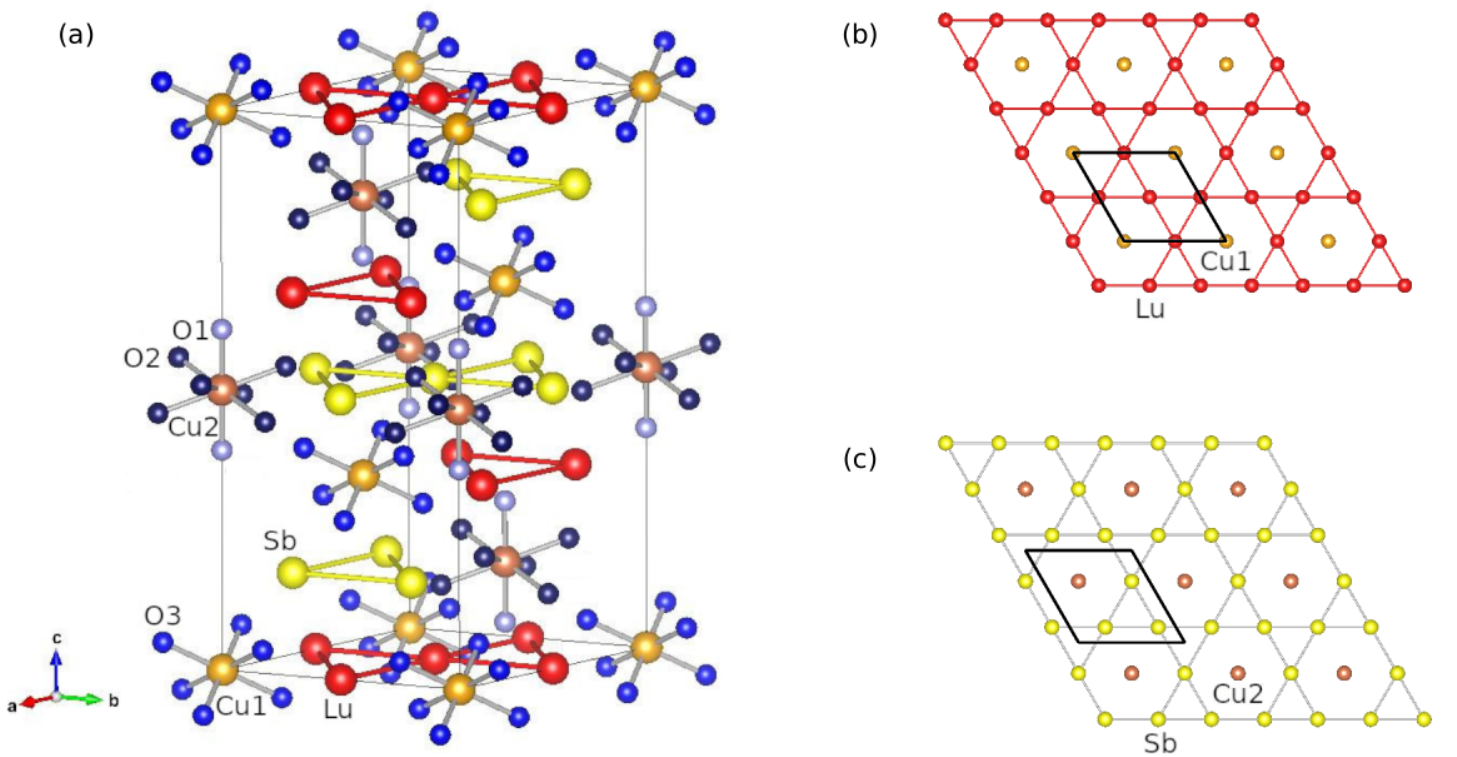} 
\caption{\label{fig:struct} Crystal structure of LCSO. (a) Unit cell, showing alternating Lu-Cu1 and Sb-Cu2 layers and oxygen coordinations of Cu1 3a and Cu2 3b sites. Inequivalent triangular Cu1 and Cu2 lattices are centered in the kagom\'e hexagons of (b) Lu layers and (c) Sb layers, respectively. The unit cells (thick black lines) are the same in both layers, showing that Cu1 sites are centered in Cu2 triangles and vice versa.}
\end{figure*}
The crystal structure was determined from diffraction data taken at room temperature using a Bruker D8 advance x-ray diffraction (XRD) spectrometer ($\lambda = 1.5418$ \AA), and the high-resolution neutron powder diffractometer (NPD) ($\lambda = 1.8846$ \AA) at the China Mianyang research reactor~\cite{ZXWX18}. XPD and NPD spectra for LCSO and LCZSO, shown in SI Sec.~S~I, exhibit narrow Bragg peaks. Parameters from Rietveld fits to the spectra are given in SI Table~S1, and the crystal unit cell is shown in Fig.~\ref{fig:struct}(a). 

\paragraph{LCSO.} Spin-1/2 Cu$^{2+}$ ions are located at the centers of the kagom\'e hexagons Cu1-Lu and Cu2-Sb, Wyckoff positions 3a and 3b respectively, and form inequivalent triangular sublattices. The 3a and 3b sites have different oxygen coordinations: 6-fold distorted octahedra and 8-fold distorted cubes, respectively. 

Our results indicate that LCSO is extraordinarily defect-free, with concentrations of ``orphan'' spins $\lesssim 10^{-3}$ and negligible Schottky magnetic defects. Static magnetism, ordered or disordered, would be expected from orphan spins, perhaps at low temperatures, but ac susceptibility and $\mu$SR experiments rule this out, the latter down to $\sim$16 mK\@. A weak Schottky specific-heat anomaly has an unobservably small effect on the entropy obtained from the specific heat.

It is important to note that when viewed along the $c$ axis, each Cu site is centered in equilateral triangles of Cu sites in adjacent layers [Figs.~\ref{fig:struct}(b) and (c)]. Section \ref{sec:symm} discusses symmetry restrictions on exchange interactions imposed by this structure. 

\begin{figure} [ht]
\includegraphics[width=0.45\textwidth]{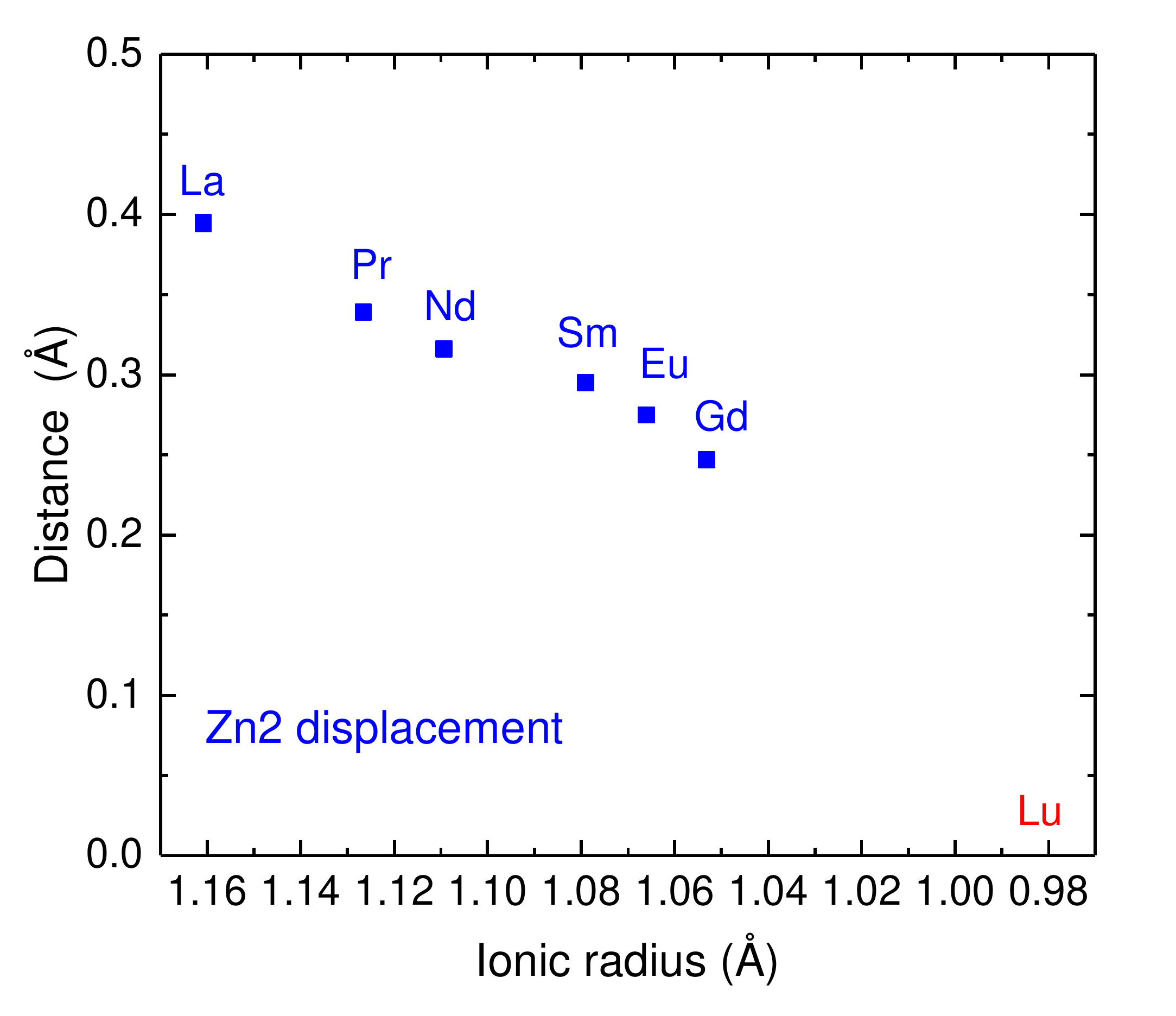} 
\caption{\label{fig:Zn2displ} Dependence of Zn2 displacement from 3b position on rare-earth ionic radius~\cite{Shan76} in \textit{RE}$_3$Zn$_2$Sb$_3$O$_{14}$. Zn2 displacements from Ref.~\onlinecite{SKC16}, ionic radii from Ref.~\onlinecite{Shan76}.}
\end{figure}

A few percent displacement of Zn2 ions from the 3b position is observed in the related compounds \textit{RE}$_3$Zn$_2$Sb$_3$O$_{14}$, \textit{RE} =  La, Pr, Nd, Sm, Eu, and Gd~\cite{SKC16}, and of Co1 ions around the 3a position in \textit{RE}$_3$Co$_2$Sb$_3$O$_{14}$, \textit{RE} =  La, Pr, Nd, Sm--Ho~\cite{LHWK14}. The offset decreases markedly with rare-earth ionic radius~\cite{SKC16, Shan76}, as shown in Fig.~\ref{fig:Zn2displ}, although it does not extrapolate unambiguously to zero for Lu.. 

However, no offset is found for Cu2 in LCSO\@. From NPD Rietveld refinements, the thermal parameter~$B$ for the Cu2 3b site $(0,0,0.5)$ in LCSO is 1.53 with no offset but significantly larger (2.26) assuming a 3\% offset ($\sim$0.2~\AA). The assumption artificially increases $B$. 

\textit{RE}/Mg site disorder has been observed in the tripod kagom\'e systems~Dy$_3$Mg$_2$Sb$_3$O$_{14}$~\cite{POHM16} and Ho$_3$Mg$_2$Sb$_3$O$_{14}$~\cite{DBPH20}. \textit{RE}$_3$Zn$_2$Sb$_3$O$_{14}$~\cite{DTLC17}, \textit{RE} = Ho, Er, and Yb, were reported to exhibit \textit{RE}/Zn disorder, but with some doubt as to the correct  model. We find no evidence for such disorder in LCSO; NPD Rietveld fits with fixed disorder result in unphysically negative thermal factors. Furthermore, disordered Cu ions on Lu 9e sites would be ``orphan'' spins, with separate contributions to the susceptibility and specific heat; these are not observed. 

\paragraph{LCZSO.} To determine whether the observed properties are specific to the 2D layers, we also synthesized Lu$_3$CuZnSb$_3$O$_{14}$ (LCZSO). We find that nonmagnetic Zn ions predominantly replace Cu2 3b ions and thus alternate with Cu1 3a layers, but the question arises whether there is any Cu/Zn site mixing. 

Cu and Zn ions have very similar coherent x-ray scattering lengths, and are hard to distinguish in XRD\@. Although their neutron scattering lengths are different (7.72 fm for Cu, 5.68 fm for Zn), we cannot obtain the Cu/Zn occupancy in LCZSO at 3a (0, 0, 0) and 3b (0, 0, ½) sites directly from NPD\@. This is because NPD intensities are different for different Cu/Zn occupancy only with odd-$l\ G$ vectors; otherwise the geometrical factor is 1 and the contribution of 3a and 3b sites to the structure factor is the sum of Cu and Zn scattering lengths. Unfortunately odd-$l$ Bragg peaks in LCZSO are too weak to determine Cu/Zn occupancy.

\begin{figure} [ht]
\includegraphics[width=0.45\textwidth]{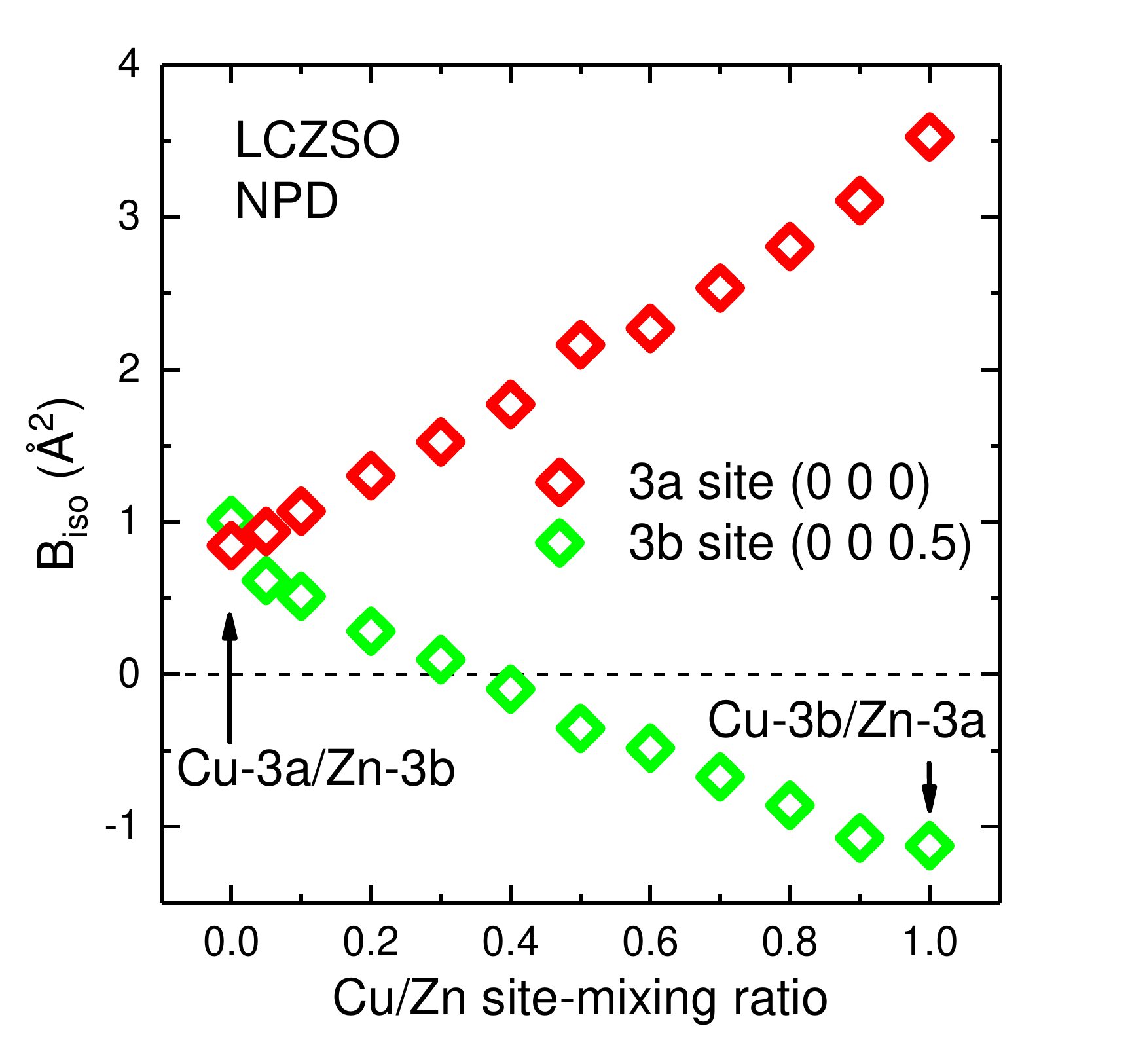} 
\caption{\label{fig:Biso} Dependence of NPD isotropic thermal parameter~$B_\mathrm{iso}$ on assumed Cu-3b/Zn-3a mixing ratio in LCZSO\@.}
\end{figure}

Instead, we use the isotropic thermal parameter~$B_\mathrm{iso}$ obtained from NPD Rietveld refinements, which is sensitive to Cu/Zn occupancy. Figure~\ref{fig:Biso} gives the dependence of $B_\mathrm{iso}$ for the two sites on the assumed Cu/Zn mixing ratio between the 3a and 3b sites. 
This ratio is 0 for 3a and 3b sites fully occupied by Cu and Zn, respectively, 1 for the opposite full occupation, and 0.5 for 0.5 Cu/0.5 Zn occupation of each site. We obtain positive $B_\mathrm{iso}$ assuming Cu 3a sites and Zn 3b sites, but negative (and therefore unphysical) $B_\mathrm{iso}$ assuming the opposite: cf.\ Fig.~\ref{fig:Biso}. 

The fits yield physically reasonable $B_\mathrm{iso}$ values (0.5--1) only for mixing ratio $\lesssim 0.1$. Refinement parameters in SI Table~S1 were obtained with this value fixed at 0.05. The 8-fold coordinated oxygen cavity centered at the 3b site is larger than the 6-fold one at the 3a site, so that $B_\mathrm{iso}$ at the 3b site should not be significantly smaller than at the 3a site. Thus the only reasonable value is for a small mixing ratio. 

We show below that a few percent impurity level is consistent with analyses of the impurity Schottky specific heat (Sec.~\ref{sec:spht}) and the low-field dc magnetic susceptibility (Sec.~\ref{sec:suscep}). We have been unable to synthesize LCZSO with less than this few percent of Cu/Zn site-interchange disorder, despite efforts with different growth protocols.

\subsection{Specific Heat and Entropy,\\ Muon Spin Relaxation} \label{sec:spht}

Specific heats were measured by the adiabatic relaxation method, using a Quantum Design Physical Property Measurement System (PPMS) equipped with a dilution refrigerator. Data were taken at temperatures between 50 mK and 300~K for LCSO and LCZSO, and 0.2~K--300~K for LZSO\@. The availability of the nonmagnetic isomorph LZSO is crucial to determining the LCSO magnetic specific heat~$C_M(T)$. 

We took special care to ensure that thermal equilibrium was achieved for the low-temperature measurements. As an example, at base temperature ($\sim$50 mK), the measurement took 70 minutes. The PPMS thermal coupling factor~\cite{[{}] [{, Sec.~4.3.3.3.}] PPMS04} between the sample and sample platform was 95\% at 100 mK and 99\% for temperatures above 0.6~K\@. Measurements were made during cooling down to base temperature as well as warming up. For measurements in a magnetic field, the sample was field cooled and then measured on warming, and also zero-field cooled, field applied at base temperature, and then measured during warming. The results were always consistent. 

A central result of this work is the observation of significantly less total magnetic entropy than the $R\ln{2}$ expected from $S=1/2$ local moments. The conclusion that the missing entropy resides at very low temperatures requires compelling evidence that it is not recovered at high temperatures. The specific heat at high temperature is, however, dominated by the lattice contribution in both LCSO and isostructural nonmagnetic LZSO\@. Below $\sim$20~K the magnetic contribution for LCSO is significant (Fig.~\ref{fig:spht}), and can be determined  accurately by simple subtraction of the LZSO specific heat. Above 20~K the difference becomes very small and difficult to measure, and it cannot be assumed that $C_\mathrm{latt}$ is the same for both compounds.

We therefore report our specific heat and entropy results from low- and high-temperature regimes in two separate sections. In the following we give results for $T \lesssim 20$~K, and discuss the regime 20--300~K in Sec.~\ref{sec:highT}.

\subsubsection{$T\lesssim 20$~K} \label{sec:lowT}

\begin{figure*} [ht]
\includegraphics[width=0.85\textwidth]{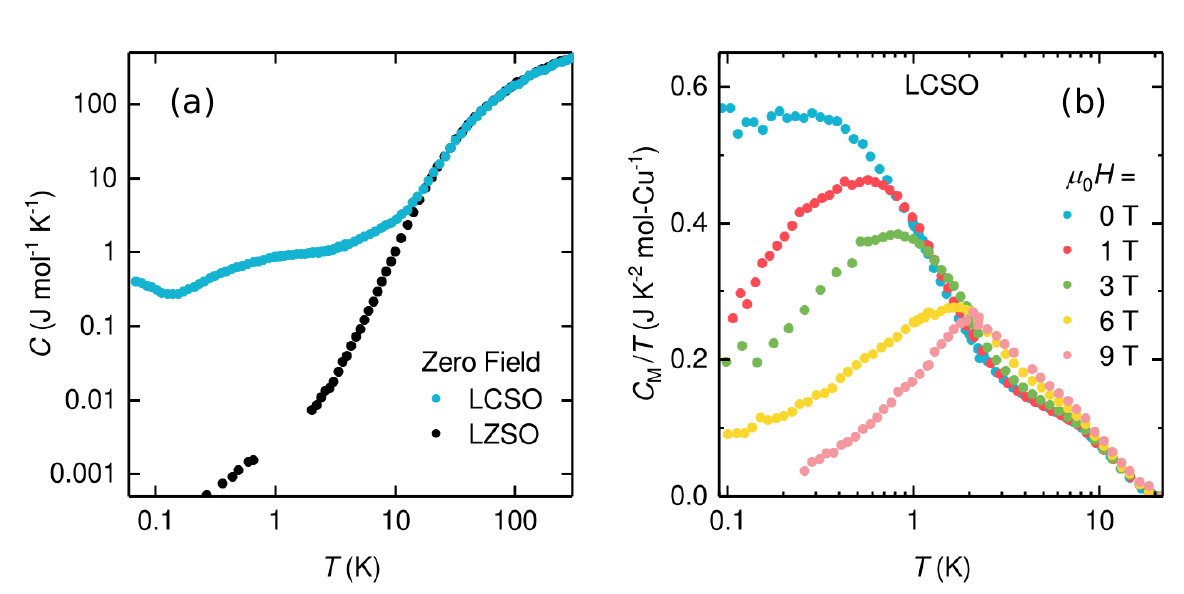}
\caption{\label{fig:spht} Specific heat of LCSO and LZSO\@. (a) Measured specific heats in zero field. (b) Intrinsic magnetic contribution $C_M(T, H)/T$ to the specific heat divided by temperature for LCSO at various magnetic fields, after subtraction of lattice, nuclear-Schottky, and impurity-Schottky contributions (see text and SI Sec.~S~II)}.
\end{figure*}

\paragraph{Specific heat and entropy.} Figure \ref{fig:spht}(a) shows the measured zero-field specific heat $C(T)$ in LCSO and LZSO below about 20~K\@. 
The latter allows an accurate subtraction of the lattice contribution $C_\mathrm{latt}(T)$. A weak ``bump'' in $C(T)$ at about 1~K and an increase below 0.2~K are observed, the magnitudes and magnetic field dependencies of which are discussed in SI Sec.~S~II\@. We attribute them to Schottky anomalies $C_\mathrm{nuc}(T)$ and $C_\mathrm{imp}(T)$, due respectively to nuclear spins and nonmagnetic impurities with excited magnetic states~\cite{He09}. Susceptibility data (Sec.~\ref{sec:suscep}) put an upper limit of ${\sim}10^{-3}$ on the concentration of impurities with magnetic ground states in LCSO. 

The intrinsic magnetic contribution $C_M(T)$ to the specific heat is then obtained by subtracting $C_\mathrm{nuc}(T)$, $C_\mathrm{imp}(T)$, and  $C_\mathrm{latt}(T)$ from the total specific heat. The various contributions to $C(T)/T$ at $H = 0$ are separately shown in SI Fig.~S2(b). 

$C_M(T,H)$ at various fields for temperatures up to 20~K is shown for LCSO in Fig.~\ref{fig:spht}(b), and for LCZSO in SI Fig.~S5(b). In LCSO $C_M(T)/T$ at $H = 0$ is constant below about 0.4~K, followed by an approximately logarithmic decrease with increasing temperature at higher temperatures. These and other features are examined in detail in Sec.~\ref{sec:2sublatt}, where the behavior is shown to be characterized by parameters close to the respective Weiss temperatures. 

\begin{figure} [hb]
\includegraphics[width=0.45\textwidth]{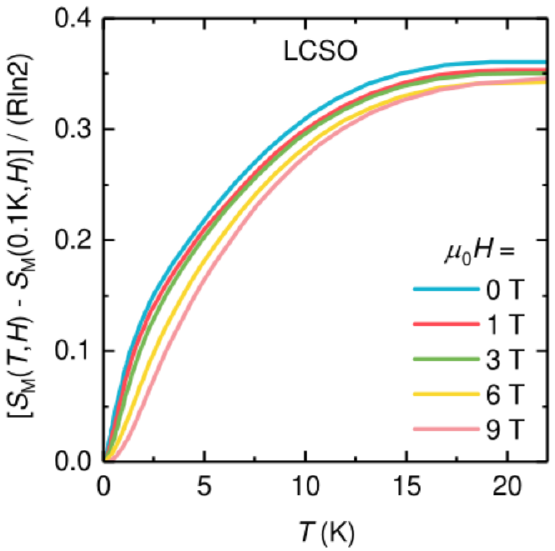}
\caption{\label{fig:ent} Change $S_M(T,H) - S_M(0.1 \text{K},H)$ in magnetic entropy, normalized to $R \ln 2$ per mol Cu, $0.1 \mathrm{K} \leq T \leq 23$~K, $0 \le \mu_0H \le 9$~T\@.}
\end{figure}

We turn next to the measurable magnetic entropy. Figure~\ref{fig:ent} shows the normalized change in magnetic entropy $[S_M(T,H){-}S_M(0.1$K$,H)]/R\ln 2$ below 20~K, calculated by integrating $C_M/T$ from 0.1~K to $T$\@. 
From the proportionality of the $\mu$SR relaxation rate to $C_M(T)/T$ from 16 mK to 0.4~K (Sec.~\ref{sec:musr} below, Fig.~\ref{fig:muSRspht}), we infer that the low-temperature constant $C_M(T)/T$ also continues down to at least 16 mK\@. At low temperatures the expected decrease of the magnitude of $S_M(T,H) - S_M(T,0)$ with increasing field~$H$ is observed (Fig.~\ref{fig:ent})\@. Although there is uncertainty in determination of $C_\mathrm{imp}(T)$ (SI Sec.~S~II~a), the entropy is affected by less than 0.5\% whether or not $C_\mathrm{imp}(T)$ is subtracted.

In LCSO for $H = 0$, $S(20\text{K}){-}S(0.1\text{K}) \approx 0.36\,k_B \ln 2$ per Cu ion, with an uncertainty of less than 2\%. As discussed in Sec.~\ref{sec:seriesexpan} in connection with Fig.~\ref{fig:hiTC+S}, for LCSO the $H = 0$ entropy at $T = 20$~K is found to be about $0.95R\ln 2$ from the two-exchange high-temperature series expansion for a triangular lattice~\cite{Sing22}. We therefore conclude that in LCSO either $\sim$60\% of the magnetic entropy ($0.95 - 0.36$) resides in the ground state or, perhaps more likely, the average $C_M/T$ below $\sim$16 mK is about 10$^3$ times the measured constant value above $\sim$100 mK.

\paragraph{Specific heat and muon spin relaxation. \label{sec:musr}} Muon spin relaxation measurements used the time-differential $\mu$SR technique \cite{Yaouanc11}, in which the evolution of the ensemble muon-spin polarization after implantation into the sample is monitored via measurements of the decay positron count-rate asymmetry $A(t)$. $\mu$SR experiments were performed down to 16 mK using the DR spectrometer on the M15 beam line at TRIUMF, Vancouver, Canada, and the DOLLY spectrometer at the Paul Scherrer Institute, Villigen, Switzerland. Samples were mounted on a silver cold-finger sample holder in the DR spectrometer, to ensure good thermal contact with the mixing chamber. $\mu$SR measurements were made from 16 mK to about 20~K in both LCSO and LCZSO (SI Sec.~S~IV)\@. Neither long-range order nor disordered spin freezing were detected in $\mu$SR asymmetry spectra down to the lowest temperatures. Appropriate functional forms of $A(t)$ were fit to the asymmetry data using the {\sc musrfit} analysis program \cite{Suter12}.

Dynamic muon spin relaxation is a direct probe of low-frequency spin dynamics \cite{Yaouanc11}. The zero-field dynamic muon spin relaxation rate $\lambda_\mathrm{ZF}(T)$ for LCSO is plotted as a function of temperature in Fig.~\ref{fig:muSRspht}. 
\begin{figure}
\includegraphics[width=\columnwidth]{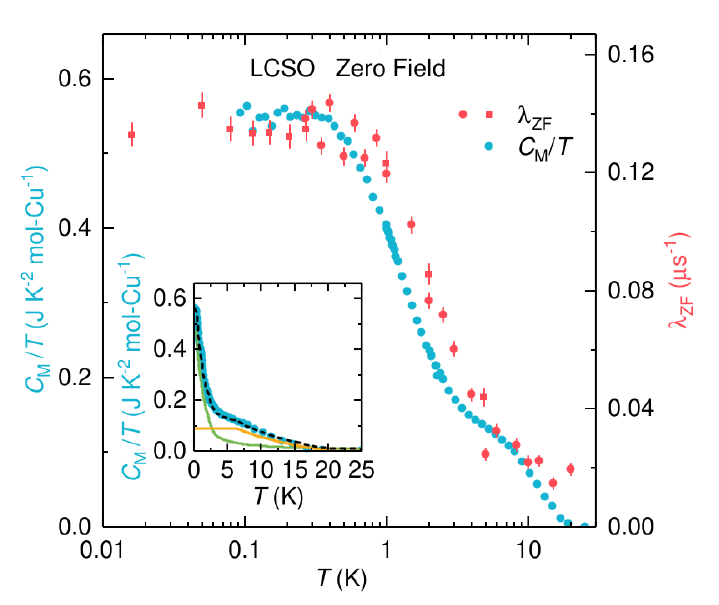}
\caption{\label{fig:muSRspht} Muon spin relaxation rate and specific heat in LCSO\@. Temperature dependencies of zero-field muon spin relaxation rate $\lambda(T)$ (red dots: data taken at PSI; red squares: data taken at TRIUMF) and $C_M(T)/T$ (blue dots) at zero field. It is remarkable that the relaxation rate tends to a constant value at low temperatures, and that it follows the temperature dependence of $C_M/T$ over the entire temperature range. Inset: separation of $C_M/T$ into contributions from the two layers (see Sec.~\ref{sec:2sublatt} below). The ratios of the low-temperature constant values and the inverses of the respective Weiss temperatures (Sec.~\ref{sec:2sublatt}) are approximately equal. The characteristic temperatures of the two logarithmic terms [Eqs.~(\ref{eq:sphtcu1}) and (\ref{eq:sphtcu2})] are also similar to the respective $\Theta_W$ values. The knee between the logarithms requires a semiclassical form (see Sec.~\ref{sec:seriesexpan}). With this fit, the {\it measured} magnetic entropy is consistent with being the same for both layers.}
\end{figure} 
Below $\sim$0.5~K $\lambda_\mathrm{ZF}(T)$ is essentially constant at ${\sim}0.14~\mu\text{s}^{-1}$, indicating persistent spin dynamics and a high density of magnetic fluctuations at low temperatures \cite{Uemura94, Ding18}. The temperature dependence of $\lambda$ closely follows that of $C_M/T$ (Fig.~\ref{fig:muSRspht}). 

In the motionally-narrowed limit
\begin{equation} \label{eq:lambda1}
\lambda(T) = \gamma_{\mu}^2 \langle B_\mathrm{loc}^2 \rangle \tau_c(T)\,, 
\end{equation}
where $\gamma_\mu = 8.5156 \times 10^4 \text{s}^{-1} \text{G}^{-1}$ is the muon gyromagnetic ratio, $\langle B_\mathrm{loc}^2\rangle$ is the mean-square fluctuation of the local magnetic fields at muon sites, and $\tau_c(T)$ is the characteristic correlation time of the local-field fluctuations. Equation (\ref{eq:lambda1}) provides a rough value of $\tau_c$ if $B_\mathrm{loc}$ is estimated as the dipolar field from a $1.85\mu_B$ Cu$^{2+}$ moment (Sec.~\ref{sec:2sublatt}) at distances to the polyhedral O$^{2-}$ ions (1.6--1.8 \AA); muon sites in transition-metal oxides are invariably near the oxygen ions. This yields $\tau_c \text{ in the range } 9\text{--}25 \times 10^{-13}$ s for LCSO and $\sim$40\% less for LCZSO\@. The relation of this estimate to the scale invariance evidenced by other properties is discussed below in Sec.~\ref{sec:invar}.

The inset to Fig.~\ref{fig:muSRspht} shows a fit to the specific-heat data for LCSO of the two-component expression
\begin{equation} \label{eq:ConTtotal}
C_M/T = C_1/T+C_2/T \,,
\end{equation}
where in units J K$^{-2}$ (mol-Cu)$^{-1}$
\begin{equation} \label{eq:sphtcu1}
C_1/T \approx \left\{ \begin{array}{cc}
 0.46 \,, & T < 0.4 \mathrm{K} \,, \\	
 0.2 \ln(4.2/T)\,, & 0.5 \mathrm{K}< T < 3 \mathrm{K} \,,\\
 0.2/T^3, & T> 3 \mathrm{K} \,,
 \end{array} \right . 
\end{equation}
and 
\begin{equation} \label{eq:sphtcu2}
C_2/T \approx \left\{ \begin{array}{cc}
 0.09 \,, & T < 7 \mathrm{K} \,, \\	
 0.096 \ln(20/T)\,, & T >8 \mathrm{K}~\,.
 \end{array}
\right . 
\end{equation}
\noindent Each component is constant at low temperatures followed by a logarithmic decrease at higher temperatures. The third term in Eq.~(\ref{eq:sphtcu1}) is the semiclassical contribution from the high-temperature series expansion [ Eq.~(\ref{eq:hiTSM})] discussed in Sect.~\ref{sec:seriesexpan}~\cite{Sing22}. The entropies of the two components up to 20~K are roughly equal. The good fit is evidence for the two-component scale-invariant ansatz discussed below in Sec.~\ref{sec:invar}. 

\subsubsection{$T\gtrsim 20$~K} \label{sec:highT}

\paragraph{Specific Heat.} \label{sec:hiTspht} It is important to measure the magnetic specific heat to high temperatures, to determine whether or not the missing magnetic entropy is generated  there. Above 20~K $C_\mathrm{nuc}(T)$ and $C_\mathrm{imp}(T)$ are negligible. $C_\mathrm{latt}(T)_\mathrm{LCSO}$ is dominant [Fig.~\ref{fig:spht}(a)], and furthermore differs slightly from $C_\mathrm{latt}(T)_\mathrm{LZSO}$. This is due to the mass difference of $\sim$0.3\% between LCSO and LCZSO and differences between Cu and Zn bonding to oxygen, which lead to a difference in $C_\mathrm{latt}(T)$ that scales with a characteristic temperature~$T^*$. The difference has a negligible effect on $C_\mathrm{diff}(T) = C(T)_\mathrm{LCSO} - C(T)_\mathrm{LZSO}$ below 20~K, but must be taken into account at higher temperatures.

We do so by taking $C_\mathrm{latt}(T/T^*)$ in the two compounds to have the same form but different $T^*~$\footnote{We thank P. A. Lee for suggesting this approach.}. We scale the LZSO temperature~$T_Z$ before subtraction by a factor~$1+\eta$, so that $C_\mathrm{diff}(T)$ is given by
\begin{equation} \label{eq:Cdiff}
C_\mathrm{diff}(T) = C_C(T) - C^{\,\prime}_Z(T) \,.
\end{equation}
Here $C_C(T)$ is the specific heat of LCSO, and $C^{\,\prime}_Z(T) = C_Z(T_Z)$ is the specific heat of LZSO with $T = (1+\eta)T_Z$. We consider $\eta$ an adjustable parameter, the calculation of which from first principles is beyond the scope of this work. Scaling the LZSO specific heat values rather than the temperature yields qualitatively similar results.

\begin{figure} [ht]
\includegraphics[width=0.45\textwidth]{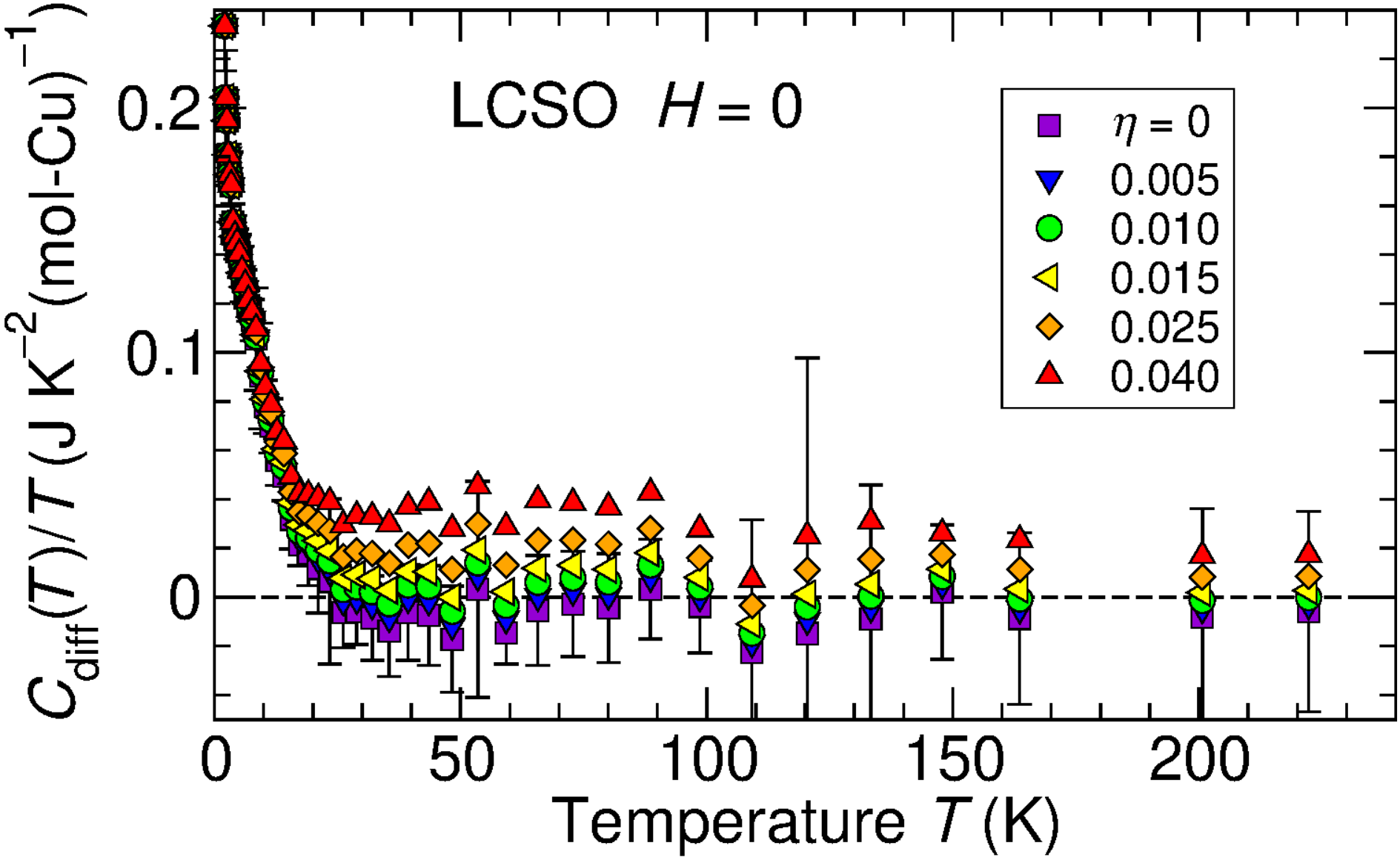}
\caption{\label{fig:Cdiff} Temperature dependence of specific heat difference $C_\mathrm{diff}(T)/T$ [Eq.~(\ref{eq:Cdiff})] for scaling indices~$\eta$ between 0 and 0.040.}
\end{figure}

Figure~\ref{fig:Cdiff} shows the temperature dependence of $C_\mathrm{diff}(T)/T$ for several values of the scaling index~$\eta$. 
For $\eta = 0$ $C_\mathrm{diff}(T)$ is negative above $\sim$20~K, i.e., $C(T)_\mathrm{LCSO} < C(T)_\mathrm{LZSO}$. This is true even if there is a magnetic contribution for LCSO\@. A negative magnetic specific heat would of course be unphysical. Positive $\eta \gtrsim 0.01$ brings $C_\mathrm{diff}$ to values $\gtrsim 0$.

We note that from Sec.~\ref{sec:spht} the missing entropy is roughly $0.6R{\ln 2}$, which corresponds to $C_M(T)/T \approx$ 0.02--0.03~J/mol~K$^2$ if spread out over 100--150~K\@. This is of the order of the error bars in Fig.~\ref{fig:Cdiff}. It is also approximately the value of $C_\mathrm{diff}/T$ for $\eta = 0.04$. Additional information is necessary to determine the correct value of $\eta$.

\begin{figure} [ht]
\includegraphics[width=0.45\textwidth]{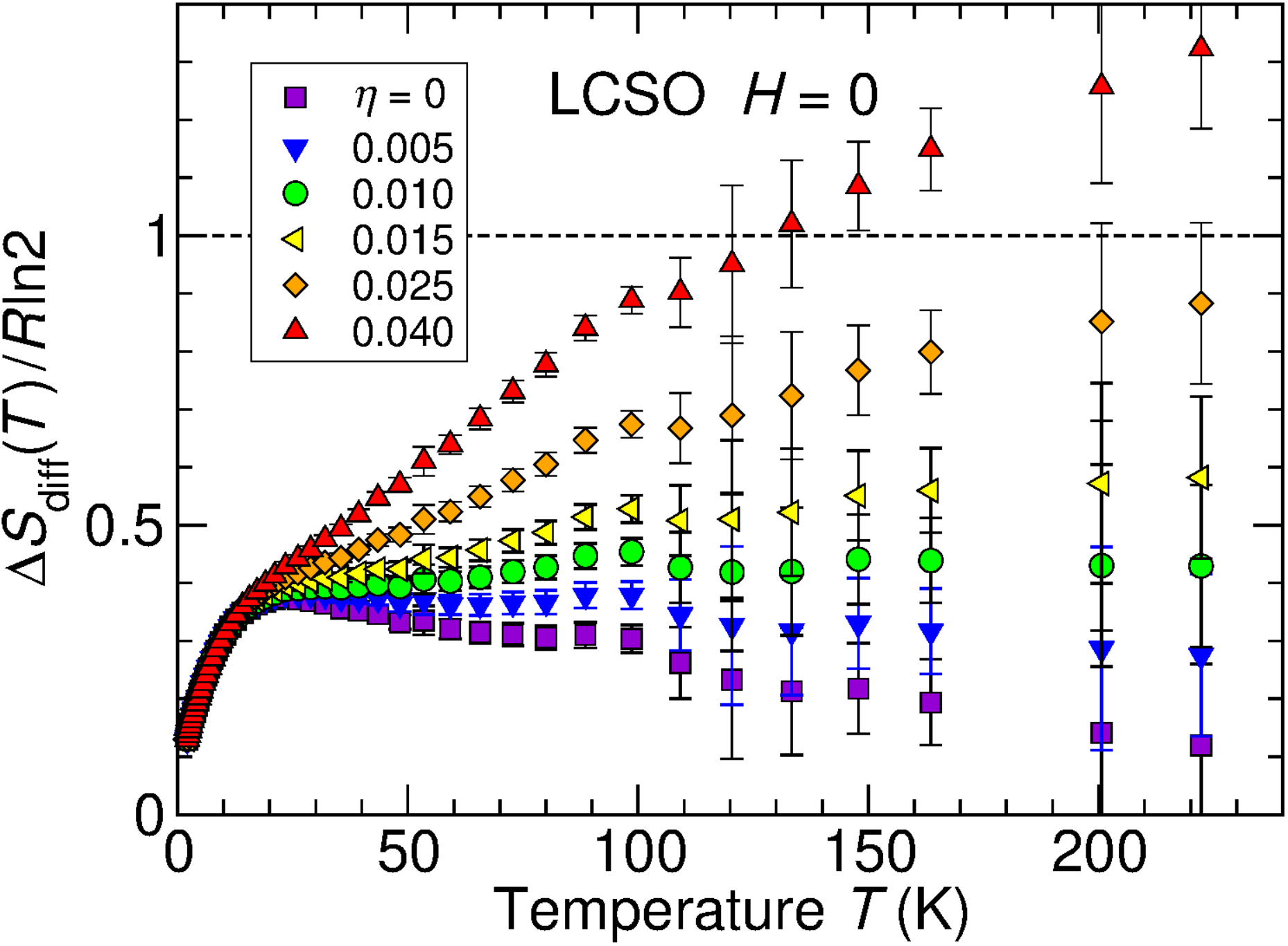}
\caption{\label{fig:Sdiff} Temperature dependence of entropy difference $\Delta S_\mathrm{diff}(T)$ for temperature scaling indices~$\eta$ between 0 and 0.040.}
\end{figure}

\paragraph{Entropy from Specific Heat.} \label{sec:hiTent} %As discussed above in Sec.~\ref{sec:spht}, the lattice specific heats of both LCSO and LZSO become very large above $\sim$20~K [Fig.~\ref{fig:spht}(a)]. 
Figure~\ref{fig:Sdiff} shows the entropy difference~$\Delta S_\mathrm{diff}(T) = S_\mathrm{diff}(T) - S_\mathrm{diff}(2\text{~K})$, calculated by integrating the data of Fig.~\ref{fig:Cdiff} between 2~K and $T$~\footnote{The error bars give the uncertainties for individual points. The scatter in the entropy is considerably less due to the integration.} and adding the measured entropy change of $0.13R\ln{2}$ between 0.1~K and 2~K (Fig.~\ref{fig:ent}).
The negative slope for $\eta = 0$ reflects the negative values of $C_\mathrm{diff}(T)$ discussed above (Fig.~\ref{fig:Cdiff}). $\Delta S_\mathrm{diff}$ is roughly constant above 20~K for $\eta = 0.01$ and above $\sim$150~K for $\eta = 0.015$. For $\eta = 0.04$ the data increase nearly linearly with increasing temperature above $\sim$25~K and pass through $R\ln{2}$ with no change.

From Fig.~\ref{fig:Sdiff} alone we cannot rule out a magnetic origin for $\Delta S_\mathrm{diff}(T)$ for $\eta \approx 0.025$. This would, however, require exchange interactions much larger than the observed Weiss temperatures from susceptibility data (Sec.~\ref{sec:suscep} below). Also, the smooth increase of $\Delta S_\mathrm{diff}(T,\eta{=}0.025)$ with increasing temperature would require a broad inhomogeneous distribution of such interactions if it were of magnetic origin. This is unlikely given the absence of evidence for defects in LCSO.

\begin{figure} [ht]
\includegraphics[width=0.45\textwidth]{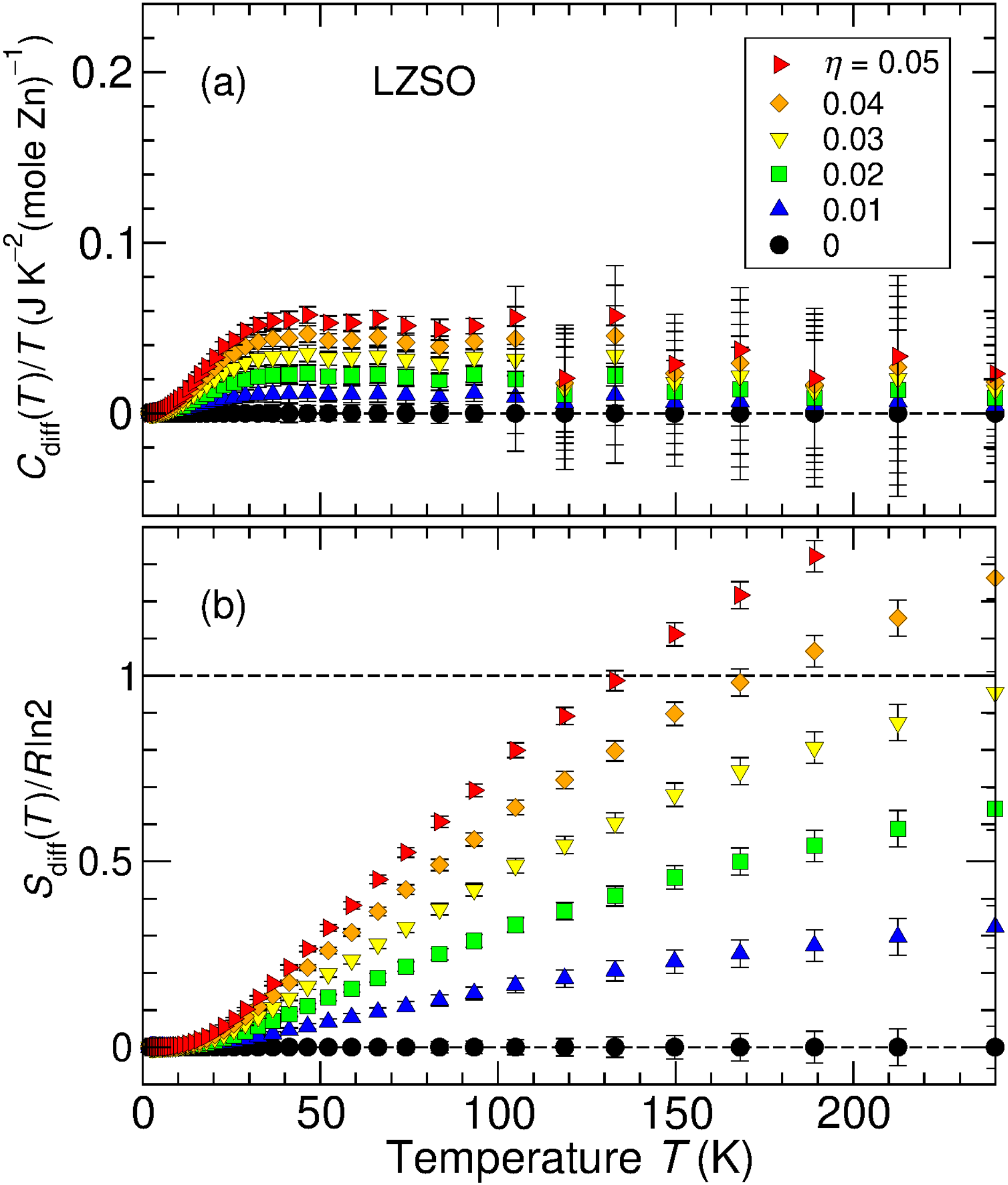}
\caption{\label{fig:CdiffLZSO} Temperature scaling of LZSO against itself. Dependence on scaling index~$\eta$ of (a)~$C_\mathrm{diff}(T)/T$ and (b)~$S_\mathrm{diff}(T)$.}
\end{figure}

The general increase of slope with increasing $\eta$ in Fig.~\ref{fig:Sdiff} is an artifact of the temperature scaling. This is demonstrated by scaling LZSO against itself, i.e., using LZSO specific heat data for both terms in Eq.~(\ref{eq:Cdiff}). Results for $C_\mathrm{diff}(T)$ and $S_\mathrm{diff}(T)$ are shown in Fig.~\ref{fig:CdiffLZSO}.
Here the increase of $S_\mathrm{diff}$ with $\eta$ is entirely due to excess subtraction, and is numerically the same as for LCSO (Fig.~\ref{fig:Sdiff}).

Furthermore, the measured exchange interactions (Sec.~\ref{sec:hiTsusc} below) are too small by an order of magnitude to be consistent with a magnetic origin of $\Delta S_\mathrm{diff}(T,\eta{=}0.025)$ in Fig.~\ref{fig:Sdiff}. The best estimate of $\eta$ is $\sim$0.01, so that there is a missing magnetic entropy of roughly $0.6R\ln{2}$. 

The ratio of atomic masses $m_\mathrm{at}(\text{LZSO})/m_\mathrm{at}(\text{LCSO})$ plays a role in determining the characteristic temperature~$T^\ast \propto m_\mathrm{at}^{-1/2}$~\cite{[{See, for example, }] [{, Chaps.~22 and 23.}]AsMe76}. This ratio is ${\sim}1.003$, which by itself would lead to $\eta \approx 0.0015$, an order of magnitude smaller than observed. This suggests that the lattice binding is significantly different in the two compounds. 

\subsection{Magnetization measurements} \label{sec:suscep}

\subsubsection{Susceptibility} 

A Quantum Design Magnetic Property Measurement System (MPMS) was used to make dc magnetic susceptibility measurements above 2~K\@. The ac magnetic susceptibility was measured for frequencies 631 Hz--10 MHz and temperatures 0.1~K--4~K in a PPMS equipped with ac susceptibility and dilution refrigerator options.

\begin{figure} [ht]
\includegraphics[width=0.45\textwidth]{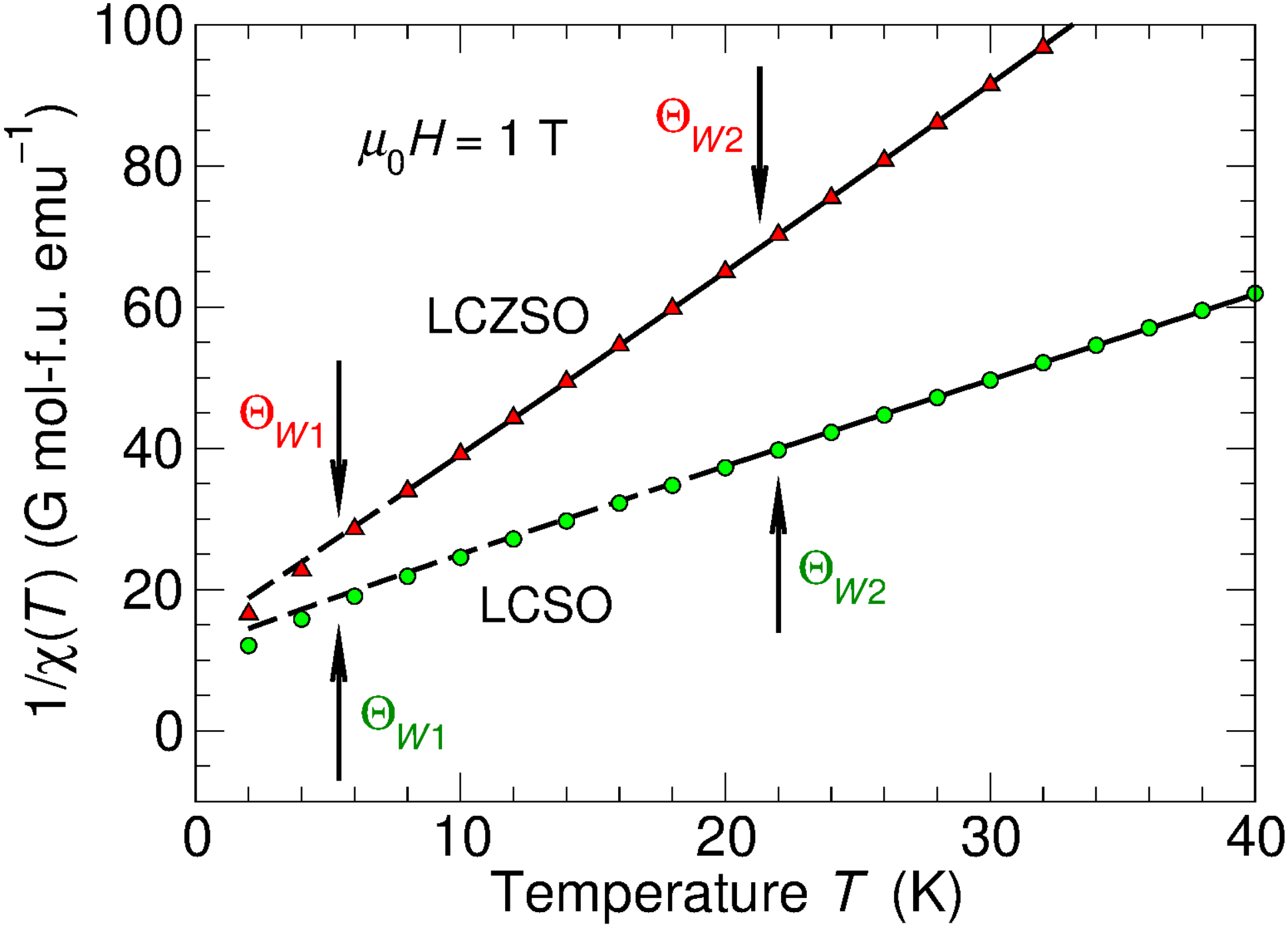}
\caption{\label{fig:CWchiinv} Temperature dependencies of inverse magnetic susceptibility in LCSO and LCZSO\@. Solid curves: fits to sum of two Curie-Weiss terms [Eqs.~(\ref{eq:CWLCSO}) and (\ref{eq:CWLCZSO})]. Dashed curves: regions of inapplicability of Eqs.~(\ref{eq:CWLCSO}) and (\ref{eq:CWLCZSO}). Arrows: Weiss temperatures $\Theta_W$.}
\end{figure}

Figure~\ref{fig:CWchiinv} shows the temperature dependencies of the inverse susceptibilities of LCSO and LCZSO\@. 
Small constant Van Vleck and diamagnetic contributions have been subtracted from the data. 

Qualitatively the data exhibit Curie-Weiss behavior; quantitative determination of exchange constants is discussed below in Sec.~\ref{sec:hiTsusc}. For both LCSO and LCZSO good fits to the data are obtained using sums of two Curie-Weiss terms:
\begin{eqnarray} 
\chi(T) & = & \chi_1(T) + \chi_2(T) \label{eq:chi1chi2} \\
& = & {\cal{C}}\left(\frac{1}{T + \Theta_{W1}} + \frac{1}{T + \Theta_{W2}}\right)\ \text{(LCSO)} \,,\label{eq:CWLCSO} \\
& = & {\cal{C}}\!\left(\frac{1 - n_2}{T + \Theta_{W1}}\,+\, \frac{n_2}{T + \Theta_{W2}}\right)\,\text{(LCZSO)} \,. \label{eq:CWLCZSO}
\end{eqnarray}
Here $\cal{C}$ is the Curie constant, $\Theta_W$'s are Weiss temperatures, and $n_2$ is the concentration of defect Cu2 ions in LCZSO\@. The fits were limited to data for temperatures $\gtrsim \Theta_W$ where the Curie-Weiss approximation is applicable, although the agreement remains good at lower temperatures.

Table~\ref{tabl:CWparams} lists the results. The effective moments and Weiss temperatures are nearly the same in the two compounds, although for LCZSO $\Theta_{W2}$ is poorly determined because of the low Cu2 concentration. 
\begin{table} [ht]
\caption{\label{tabl:CWparams} Parameters from fits of Eqs.~(\ref{eq:CWLCSO}) and (\ref{eq:CWLCZSO}) to susceptibility data (Fig.~\ref{fig:CWchiinv}).}
\begin{ruledtabular}
\begin{tabular}{lcc}
 & LCSO & LCZSO \\
\colrule
${\cal{C}}$ [emu K G$^{-1}$ (mol f.u.)$^{-1}$]  & 0.426(1) & 0.4190(6)  \\
Effective moment from ${\cal{C}}$ ($\mu_B$) & 1.846(2) & 1.830(1)  \\
$\Theta_{W1}$ (K) & 6.7(6) & 5.4(3) \\
$\Theta_{W2}$ (K) & 18.9(1.2) & 22(13) \\
LCZSO Cu2 concentration $n_2$ & -- & 0.07(6)
\end{tabular}
\end{ruledtabular}
\end{table}
The value of $n_2$ for LCZSO is in agreement with that for site-interchange Cu2 ions on the 3d Zn layers from the structure determination (Sec.~\ref{sec:struct}). These results identify terms 1 and 2 in Eqs.~(\ref{eq:chi1chi2})--(\ref{eq:CWLCZSO}) with Cu1 3a and Cu2 3b sites, respectively, since Cu ions predominantly occupy the 3a site in LCZSO.

\begin{figure} [ht] 
\includegraphics[width=0.45\textwidth]{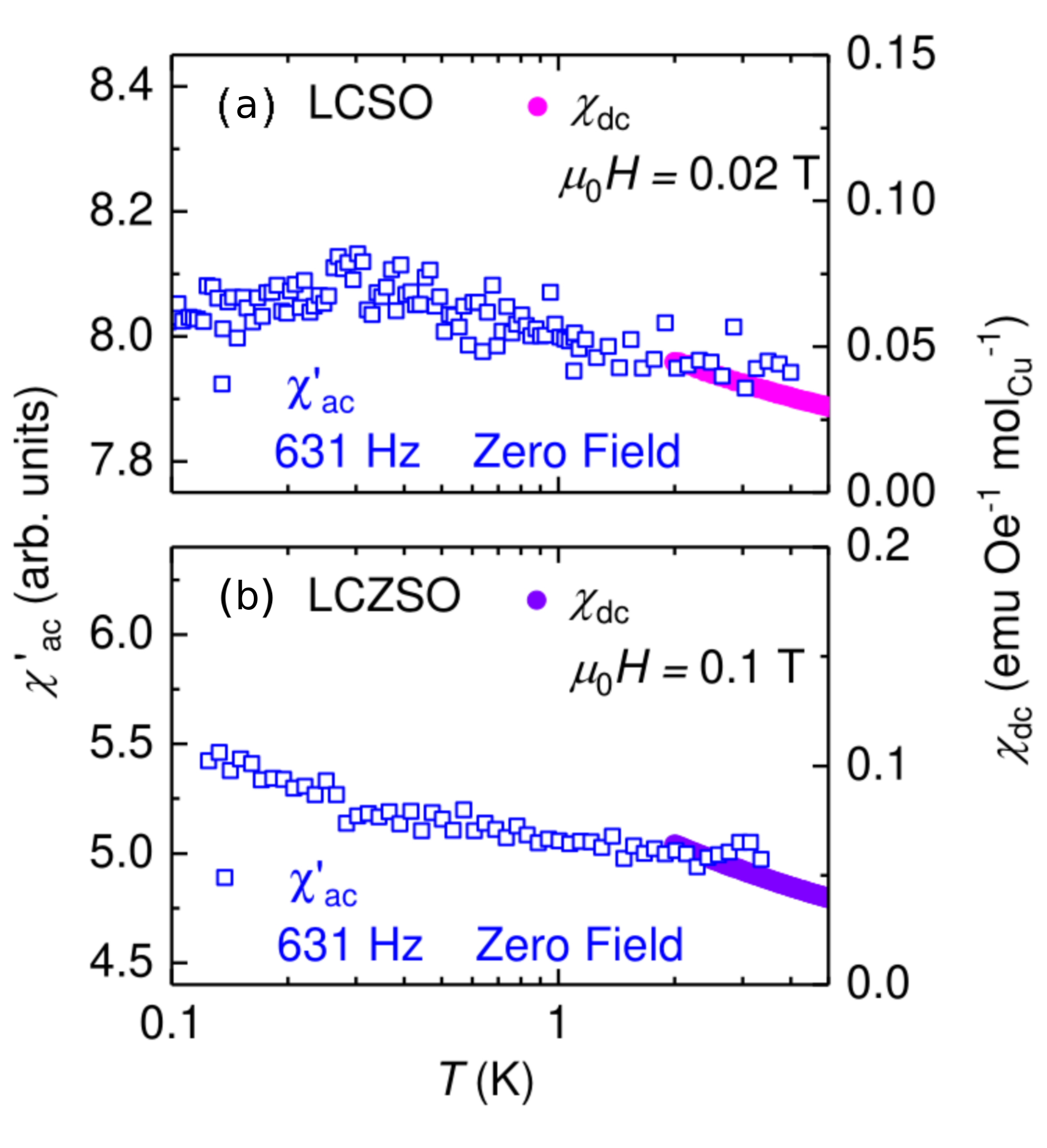}
\caption{\label{fig:chiac} Temperature dependencies of the real part $\chi'_\mathrm{ac}$ of the ac susceptibility in (a) LCSO and (b) LCZSO, 0.1~K $\le T \le 4$~K\@. Scales are adjusted to match $\chi_\mathrm{dc}$ above 2~K\@. Similar temperature dependencies were found for frequencies up to 10 kHz.}
\end{figure}

Any additional magnetic impurity concentration~$n_\mathrm{imp}$ was estimated by adding a Curie law ($\Theta_W \approx 0$, appropriate to nearly-free impurity spins) to the fit function. This yields upper limits $n_\mathrm{imp} \sim 5 \times 10^{-3}$ for both compounds. Stronger estimates of impurity concentrations are obtained from the real part $\chi_\mathrm{ac}'$ of the ac susceptibility measured over the temperature range 0.1~K--3~K, shown in Figs.~\ref{fig:chiac}(a) (LCSO) and \ref{fig:chiac}(b) (LCZSO)\@. 
By ascribing all the temperature dependence of $\chi_\mathrm{ac}'$ in this temperature range to a Curie law, we find upper limits to the concentrations of free-spin magnetic impurities of about $10^{-3}$ in LCSO and about twice that value in LCZSO\@. 
 
A broad maximum is observed at $\sim$0.3~K for LCSO [Fig.~\ref{fig:chiac}(a)]. No frequency dependence was found up to 10 kHz (data not shown) and there is no signature of spin freezing in the $\mu$SR data (Sec.~\ref{sec:musr}), so that a spin-glass transition is unlikely. The maximum is not understood, but it is interesting that the crossover of $C(T)/T$ and the muon spin relaxation ($\mu$SR) rate $\lambda_\mathrm{ZF}(T)$ from constant to logarithmic temperature dependence (Sec.~\ref{sec:musr}, Fig.~\ref{fig:muSRspht}) occurs in this region. The effect is small; the area under the maximum is only a few percent of the total area under the susceptibility up to 3~K\@.

\begin{figure} [ht]
\includegraphics[width=0.45\textwidth]{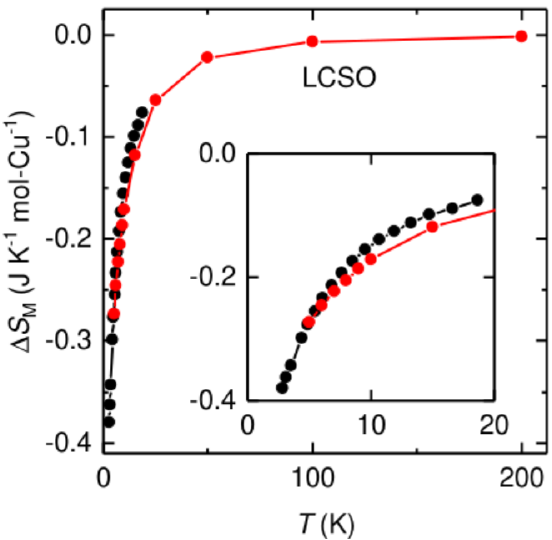}
\caption{\label{fig:magent} Entropy of LCSO\@. Black symbols: from specific heat measurements. Red symbols: from measurements of magnetization and the Maxwell relation~$(\partial S/\partial H)_T = (\partial M/\partial T)_H$. The lost magnetic entropy is fully recovered at high temperatures, indicating that the missing entropy is independent of the applied field.}
\end{figure}

\subsubsection{Entropy from magnetization.} 

We determine characteristics of the entropy in a magnetic field up to 200~K by an alternate %more accurate 
method, which also gives an estimate of the accuracy of the subtraction procedure below 20~K\@. In this method, the magnetization $M(H,T)$ is measured from 4~K to 300~K at various fields [SI Figs.~S3(a) and (c) for LCSO and LCZSO, respectively]. $(\partial M/\partial T)_H$ are shown in SI Figs.~3(b) and 3(d), respectively. We then use the Maxwell relation
\begin{equation} \label{eq:magent}
\left(\frac{\partial S}{\partial H}\right)_{\!T} = \left(\frac{\partial M}{\partial T}\right)_{\!H}, 
\end{equation}
and integrate $(\partial S/\partial H)_T$ to give the change $\Delta S(T,H)$ in entropy due to the magnetic field as a function of temperature. The results are displayed as red points in Fig.~\ref{fig:magent}. At high temperatures $\Delta S(T,H) \to 0$, indicating approach of the entropy to a field-independent value. 

The results from direct determination by subtraction below 20~K are shown in Fig.~\ref{fig:magent} as black points. The agreement provides a quantitative measure of the consistency between the results of these quite different techniques, see Fig.~\ref{fig:magent} and SI Fig.~S5(d). To get a measure of the consistency of results obtained by these quite different methods, we note that the standard deviation of the red and black points in LCSO is $\sim$0.02. 

The results in Fig.~\ref{fig:magent} show that the \textit{available} entropy loss due to magnetic fields at low temperatures is monotonically reduced to zero at higher temperatures. However, the \textit{missing} entropy is field independent up to 9 T over the whole temperature range. From this behavior at $g\mu_B H \leq k_BT$ it follows that the missing entropy is due to purely singlet excitations. Since it is unaffected even for $g\mu_B H \gtrsim k_B\Theta_W$ (Sec.~\ref{sec:2sublatt}), local mutually non-interacting singlet states are also ruled out because they would be replaced by multiplet states favored by magnetic polarization. There is no history dependence or hysteresis (Sec.~\ref{sec:spht}), so that the phenomena do not appear to be due to metastable singlet states. The singlet states must be collective and therefore non-local, with collective barriers against thermally-excited magnetic states.

\vspace{20pt} \section{DISCUSSION} \label{sec:disc}
~
\vspace{-40pt} \subsection{High-temperature series expansions for a triangular lattice} \label{sec:seriesexpan}

Magnetic entropy at high temperatures could arise from stronger exchange interactions than implied by Weiss temperatures~$\Theta_W$ from the Curie-Weiss fits of Sec.~\ref{sec:suscep} (Table~\ref{tabl:CWparams}), since nearest-neighbor antiferromagnetic (AFM) and 2nd-nearest-neighbor ferromagnetic (FM) exchange tend to cancel in $\Theta_W$. To examine this situation we use high-temperature ``semiclassical'' truncated series expansions for the magnetic susceptibility~$\chi(T)$ and entropy~$S_M(T)$ of the Heisenberg model on a triangular lattice. These are available for nearest-neighbor  interactions~\cite{ESY93}, and also for nearest- and 2nd-nearest-neighbor interactions~\cite{Sing22} with exchange interactions~$J_1$ and $J_2$, respectively. 

The first three terms of the $J_1$-$J_2$ expansions are
\begin{eqnarray}
\chi(T)T/{\cal C} & = & 1 - \frac{3}{2}(J_1 + J_2)/T  \nonumber \\
& & + \frac{3}{2}(J_1^2 + J_2^2 + 3J_1J_2)/T^2 + \cdots \,, \quad \label{eq:hiTchi}
\end{eqnarray}
for the susceptibility, where $\cal C$ is the Curie constant, and
\begin{eqnarray}
S_M(T)/R & = & \ln{2} - \frac{9}{32}\left(J_1^2 + J_2^2\right)/T^2 \nonumber \\
& & + \frac{9}{96}\left(J_1^3 + J_2^3 +6J_1^2J_2\right)/T^3 + \cdots \,, \quad\label{eq:hiTSM} 
\end{eqnarray}
for the entropy, where $R$ is the gas constant. The accuracy of the truncated series is limited to temperatures above roughly twice the larger exchange~\cite{Sing22}.

\subsubsection{Susceptibility} \label{sec:hiTsusc}

We first fit high-temperature series expansions to the susceptibility data (Sec.~\ref{sec:suscep}). For LCSO the fit function consists of two terms of the form of Eq.~(\ref{eq:hiTchi}), one for each of the two layers. A fit with all four exchange parameters free is not possible, because of the inherent ambiguity in a sum of two terms of the same form~\footnote{In a function of the form $y = Ax + Bx$ the parameters $A$ and $B$ cannot be fit for separately, only the sum~$A+B$.}. We therefore use results from fits of Eq.~(\ref{eq:hiTchi}) to the LCZSO susceptibility, which is dominated by the contribution of the Cu1 layer (Sec.~\ref{sec:2sublatt}). For the LCSO fit the Cu1 layer exchanges were fixed at the LCZSO values, on the grounds that the properties of this layer are not very different in the two compounds (cf.\ Sec.~\ref{sec:suscep}).

\begin{figure} [ht] 
\includegraphics[width=0.45\textwidth]{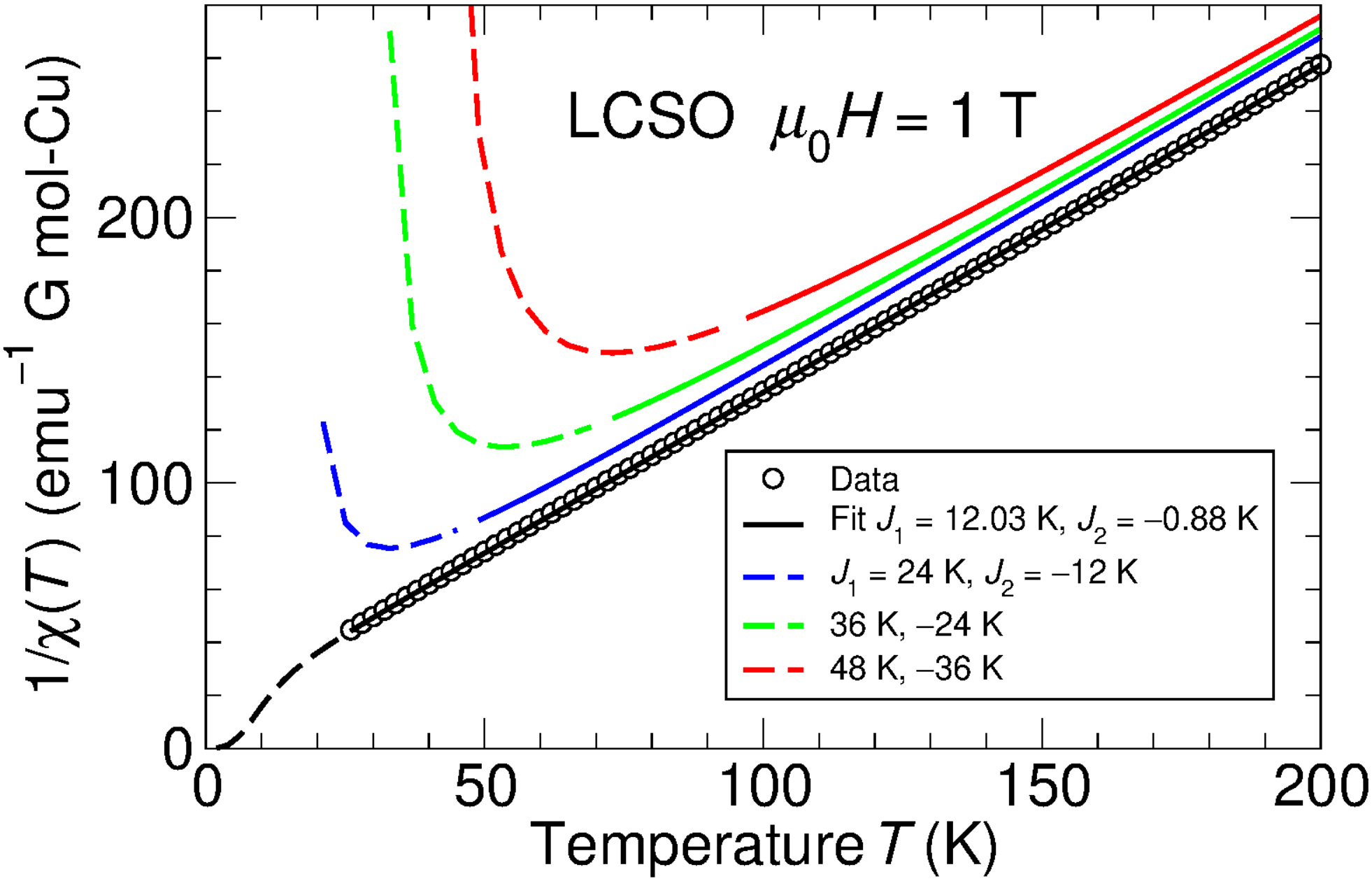}
\caption{\label{fig:highTsusc} Comparison of LCSO magnetic susceptibility with truncated high-temperature series expansions [Eq.~(\ref{eq:hiTchi})]. Black curve: fit of Eq.~(\ref{eq:hiTchi}) to data for $T > 25$~K\@. Other curves: series expansions for larger exchange values. Dashed curves: regions of inapplicability of Eq.~(\ref{eq:hiTchi}).}
\end{figure}

Figure~\ref{fig:highTsusc} shows the fit to data for $T > 25$~K, roughly twice the fit value of $J_1$.
$J_2$ is negative (FM) and small. Values of the fit parameters are given in Table~\ref{tabl:exch}.
\begin{table} [ht]
\caption{\label{tabl:exch} Curie constant~$\cal{C}$ and exchange parameters in LCSO from fits of two-exchange high-temperature series expansions to susceptibility data (Fig.~\ref{fig:highTsusc}).}
\begin{ruledtabular}
\begin{tabular}{lcc}
 & LCSO\footnote{From sum of two terms of form of Eq.~(\ref{eq:hiTchi}); see text.} & LCZSO\footnote{From Eq.~(\ref{eq:hiTchi}).} \\
\colrule
${\cal{C}}$ [emu K G$^{-1}$ (mol f.u.)$^{-1}$]  & 0.4203(9) & 0.4170(2)  \\
$J_{1}$ (K) & 12.03(12) & 3.54(6) \\
$J_{2}$ (K) & $-$0.88(9) & 0.39(8) \\
\end{tabular}
\end{ruledtabular}
\end{table}

For comparison a number of curves are shown for various larger $J_1$ and $J_2$, such that $J_1 + J_2$ in the leading term in Eq.~(\ref{eq:hiTchi}) is roughly constant. There is no agreement. This is strong evidence that exchange in LCSO is weak and does not produce interaction phenomena at higher temperatures.

\begin{figure} [ht] 
\includegraphics[width=0.45\textwidth]{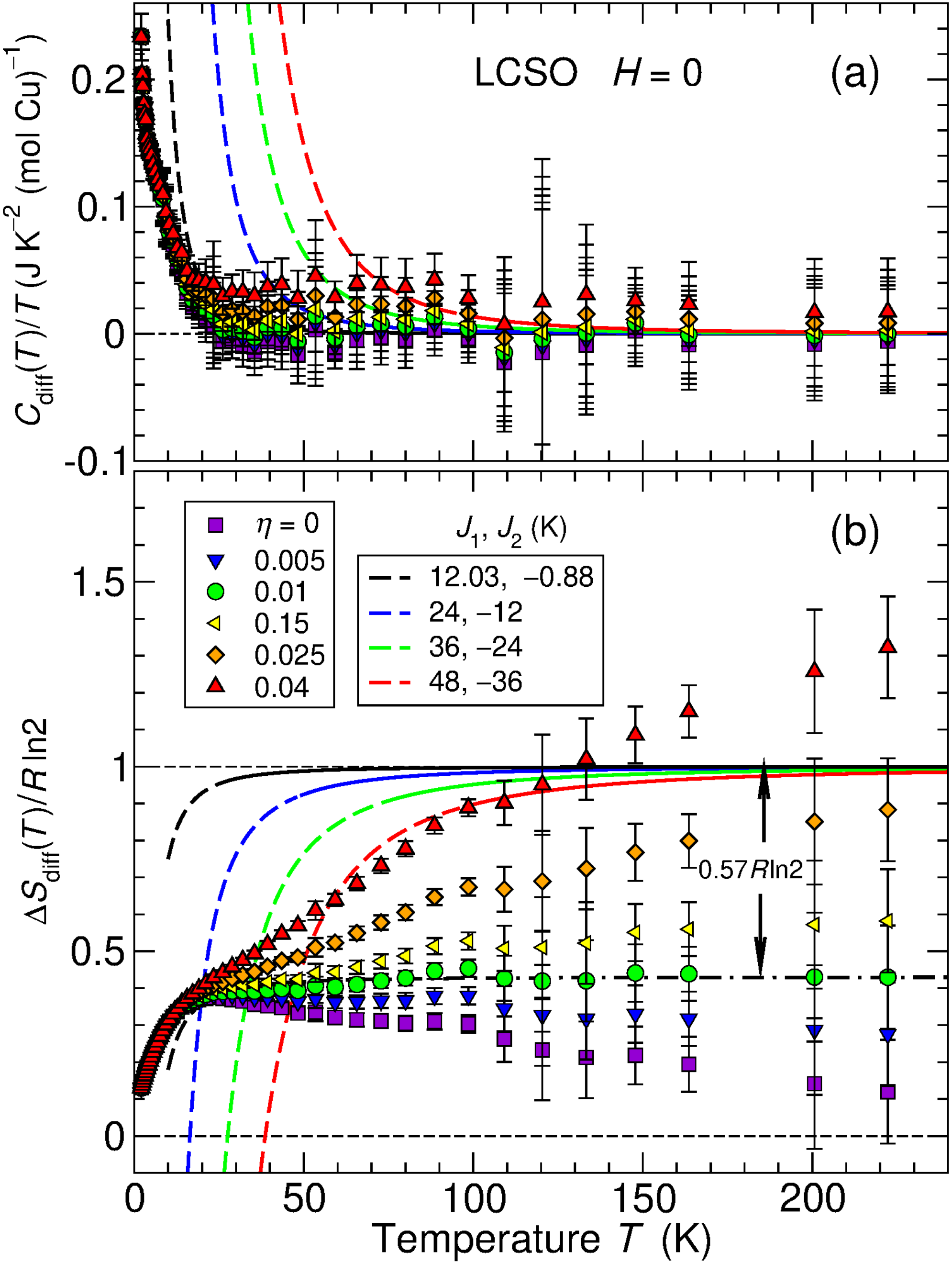}
\caption{\label{fig:hiTC+S} (a) Specific heat difference $C_\mathrm{diff}(T)/T$ and (b) entropy difference $\Delta S_\mathrm{diff}(T)$ from Figs.~\ref{fig:Cdiff} and \ref{fig:Sdiff}, compared to truncated high-temperature expansions [Eq.~(\ref{eq:hiTSM}) and its derivative]. Dash-dotted curve: entropy expansion for $J_1 = 14$~K, $J_2 = -2$~K with $0.57R\ln{2}$ subtracted. Dashed curves: regions of inapplicability of Eq.~(\ref{eq:hiTSM}).}
\end{figure} 

\subsubsection{Specific heat and entropy} \label{sec:hiTsphtent}

The high-temperature specific heat and entropy differences reported in Sect.~\ref{sec:highT} are replotted in Fig.~\ref{fig:hiTC+S}, together with high-temperature expansions with exchange values from Table~\ref{tabl:exch} and Fig.~\ref{fig:highTsusc}. Our purpose is to determine the correct value of the scaling parameter~$\eta$. From Fig.~\ref{fig:hiTC+S}(b) a candidate might be $\eta = 0.04$, which between 50~K and 150~K agrees roughly with the expansion for $J_1 = 48$~K, $J_2 = -36$~K\@. However, results of fits to susceptibility data (Table~\ref{tabl:exch}) rule out such large exchange values. In any case the data pass smoothly through $R\ln{2}$, which is not possible for magnetic entropy.

For exchange values compatible with susceptibility fits, the entropy series expansions are constant down to low temperatures. The data for $\eta = 0.01$ (green circles) agree with this property, as shown in Fig.~\ref{fig:hiTC+S}(b). This is strong evidence that for those data $\Delta S_\mathrm{diff}(T) = \Delta S_M(T)$, the magnetic entropy. Then the series expansion agrees very well with the data after subtracting a missing entropy of $0.57 R\ln{2}$ [dash-dotted curve in Fig.~\ref{fig:hiTC+S}(b)].

\subsection{\boldmath Coordination of Cu$^{2+}$ ions,\\ symmetry of the orbitals} \label{sec:symm}

We consider the symmetry of the Cu orbitals where the spins reside and the exchange path. We show by a symmetry argument that the magnetic interactions are two-dimensional. Accordingly, all data discussed below in Sec.~IV can be separated into distinct components associated with the two lattice planes, each characterized by a separate parameter (the Weiss temperature). Experimental evidence for this conclusion is discussed in Secs.~\ref{sec:suscep} and \ref{sec:2sublatt}.

The O$^{2-}$ coordination of the Cu1 3a site is a trigonally-distorted octahedron, and that of the Cu2 3b site is a trigonally-distorted cube (SI Sec.~I). The 3a-site distorted octahedron is oriented with the $c$ axis passing through the centers of two parallel triangular faces. The [111] corners of the 3b-site distorted cube are compressed along the $c$ axis. Thus there is 3-fold rotation symmetry around the $c$ axis at both sites, as dictated by the crystal symmetry.

A general symmetry argument can be made that the inter-plane coupling vanishes despite the short interlayer distance [Fig.~\ref{fig:struct}(a)]. It relies on two facts: (1) there are common three-fold axes for the two planes, each of which passes through a Cu ion in one layer and the center of an equilateral triangle of three Cu ions in the adjacent layer [Figs.~\ref{fig:struct}(b) and (c)]; and (2) the relevant Cu orbitals in the two layers are different and orthogonal.

We recall standard results from crystal-field theory \cite{[{See, e.g., }]Burns93}: the highest-energy $3d$ states, which the hole occupies in Cu$^{2+}$ ions, are the two $e_g$ states in octahedral coordination and the three $t_{2d}$ states in cubic coordination. Consider a centered Cu2 ion and the equilateral triangle of Cu1 ions in the next layer. Let $F_1(\theta,\phi)$ and $F_2(\theta,\phi)$ be the Wannier orbitals constructed from the three $e_{g}$-derived Cu1 wave functions and their associated oxygen polyhedra on the triangle, and $F_i(\theta,\phi),~i = 3,4,5$, the $t_{2g}$-derived Cu2 wave functions in the distorted cube. The angles $\theta$ and $\phi$ are defined by the common three-fold axis of symmetry. All five $F_i$ are second-order functions of angle. 

Then the most general wave functions, which in general are nondegenerate, projected to the base of the triangle are 
\begin{equation}
\psi_1 = (1/N1)[F_1(\theta, \phi)\cos 3\phi + F_2(\theta,\phi) \sin 3\phi]
\end{equation} 
for the Cu1 triangle, and
\begin{eqnarray}
\psi_2 & = & (1/N2)[F_3(\theta,\phi) \nonumber \\
 & & + F_4(\theta,\phi) \cos 3\phi + F_5(\theta,\phi) \sin 3\phi]
\end{eqnarray}
for the Cu2 ion.

Because of the orthogonality of the $F_i$ and the fact that they do not contain any odd polynomials like $\sin \phi$, $\sin 3 \phi$, etc., $\psi_1$ and $\psi_2$ are orthogonal. Since all potentials have three-fold symmetry, any matrix elements of $\psi_1$ and $\psi 2$ with any such potential is also zero.

This suggests that in LCSO, even though the interlayer distance between Cu sites is shorter than the intralayer distance, superexchange interaction between layers is likely to be absent. Then LCSO (and LCZSO) are two-dimensional as far as the magnetic interactions between the spin-1/2 Cu$^{2+}$ ions are concerned. As a consequence LCSO has similar properties per Cu ion as LCZSO, where Cu ions only occupy the Lu layer. However, more evidence, e.g, from measurements of crystal-field levels in single crystals, would be desirable to substantiate this conclusion.

As may be seen in Fig.~\ref{fig:struct}(a), the superexchange path between nearest-neighbor Cu ions in a layer is via their oxygen polyhedra and the intervening Lu or Sb ion. This long path makes the nearest-neighbor exchange energy small. It also means that an estimate of it (let alone 2nd-nearest-neighbor exchange or triple exchange among three Cu's) will be very hard to obtain from a calculation. This is fortunately not necessary for the purposes of the present paper.

\subsection{Two sublattices: specific heat and susceptibility} \label{sec:2sublatt}
 
Inequivalent Cu1 3a and Cu2 3b sites in LCSO form two separate triangular sublattices coordinated by Lu and Sb ions, respectively. Here we present evidence for two additive components in the magnetic specific heat and the magnetic susceptibility that are due to the two sublattices. This is strong evidence that their properties are two-dimensional, with little interlayer interaction, in agreement with the preceding discussion (Sec.~\ref{sec:symm}).

\paragraph{Specific heat.}  In Sec.~\ref{sec:musr}, we show in the inset to Fig.~\ref{fig:muSRspht} that the measured $C_M(T)/T$ in LCSO can be separated into two components. The ratio of their low-temperature constant values is approximately inversely proportional to their respective values of $\Theta_W$, and they both decrease logarithmically approximately as the corresponding $\ln(\Theta_{W}/T)$\@. The integrated entropies are about the same for the two layers. These forms only pertain for the ``quantum region'' below the respective $\Theta_W$'s. The knee region between the two logarithms requires fit to the semiclassical region (Sec.~\ref{sec:seriesexpan}). 

A similar decomposition for LCZSO is discussed in SI Sec.~S~IV\@. There it is shown that, even with only a few percent site interchanges of Cu and Zn (see SI Sec.~S~I and below), the ratio of the entropies of the two layers is no smaller than 30\%. This emphasizes the importance of defect-free samples for the study of intrinsic properties of spin liquids.

\paragraph{Susceptibility.} \label{par:suscept} In Sec.~\ref{sec:suscep} we show results of dc and ac susceptibility measurements down to 0.1~K (Figs.~\ref{fig:CWchiinv} and \ref{fig:chiac}), estimate the magnetic impurity concentrations, and discuss two-component Curie-Weiss fits for both LCSO and LCZSO\@. We find magnitudes of the components (equal in LCSO, approximately in the ratio 93:7 in LCZSO) and values of $\Theta_W$ (Table~\ref{tabl:CWparams}) that are strong evidence for two weakly-interacting sublattices associated with the inequivalent layers. 

It is important to note that the characteristic temperatures in the logarithms in the specific heat [Eqs.~(\ref{eq:sphtcu1}) and (\ref{eq:sphtcu2})] are close to the Weiss temperatures or characteristic exchange energies (Table~\ref{tabl:CWparams}), and that the ratio of the constant specific heats at low temperature is approximately the inverse ratio of the $\Theta_W$'s \footnote{The break points in Eqs.~(\ref{eq:sphtcu1}) and (\ref{eq:sphtcu2}), which were chosen for best fit, do not scale with the $\Theta_W$'s.}. This is evidence that in LCSO the two Cu layers are contributing equally to both quantities.

In LCZSO there is considerable uncertainty in the Cu2 concentration from the Curie-Weiss fit (Sec.~\ref{sec:suscep}, Table~\ref{tabl:CWparams}). Nevertheless, the results are consistent with the evidence from the structure analysis for a few percent Cu occupation of the 3b (Zn) sublattice. The near equality of $\Theta_{W1}$ values for the two compounds (Table~\ref{tabl:CWparams}) is evidence that $\chi_1$ and $\chi_2$ in Eq.~(\ref{eq:chi1chi2}) are the contributions of the Cu1 and Cu2 layers, respectively. Our observations support the ansatz of additive effects of the two inequivalent Cu sublattices.

\subsection{Scale invariance} \label{sec:invar}

The properties reported here can be used to specify some features of the frequency-dependent correlation functions that a fundamental theory might provide. We consider only the pure limit and the experimental results for $H = 0$. We show that the observed specific heat and muon spin relaxation rate follow if there are magnetic fluctuations with a \emph{local} density of states $\mathcal{A}_M(\omega,T)$ of a specific scale-invariant form. The ground-state entropy requires a more singular form $\mathcal{A}_0(\omega,T)$. 

We write
\begin{align} \label{eq:Aloc}
\mathcal{A}_\mathrm{loc} (\omega,T) & \equiv \sum_\mathbf{q} \mathcal{A}(\mathbf{q},\omega,T) \delta(\omega - \omega_\mathbf{q}) \nonumber \\
& = \mathcal{A}_0(\omega,T) + \mathcal{A}_M(\omega,T) \,,
\end{align}
where
 \begin{equation} \label{eq:A0}
\mathcal{A}_0(\omega,T) = S_0\frac{\omega}{\omega^2 + T^2} e^{-\omega/T}\,, \\
\end{equation}
and
\begin{align} \label{eq:AM}
\mathcal{A}_M(\omega,T) & = \gamma_M \left\{ \begin{array}{ll}\displaystyle \frac{\omega}{T} \left| \ln \frac{T + T_x'}{T_x}\right|\,, & \omega \ll T\,, \\
\displaystyle \left|\ln\frac{\omega}{T+T_x}\right|\,, & T \lesssim \omega \lesssim T_x\,.
\end{array} \right. 
\end{align}
Here $T_x \approx \Theta_W$, and the experiments yield $T_x' \approx T_x/4$. The quantum numbers $\mathbf{q}$ of the fluctuations are different for the two contributions. 

As shown below, $\mathcal{A}_M(\omega,T)$ provides the measured specific heat $C_M/T \approx (\gamma_M k_B)$ for $T \ll T_x$%, and the $\mu$SR relaxation rate, which requires magnetic field fluctuations
. The local density of states of singlet excitations $\mathcal{A}_0(\omega,T)$ can easily be modified if later experiments reveal a gigantic peak in $C_M/T$ at very low temperatures. $S_0$ and $\gamma_M T$ with a cutoff $T_x$ of order the Weiss temperatures, are related through the sum rule that the total entropy is $k_B \ln 2$ per spin at high temperatures. 

The fluctuations are assumed to obey Bose-Einstein statistics with zero chemical potential. The reason this works is that even with zero chemical potential, as can be easily calculated, the number of excitations remains independent of temperature with the choice of the singular density of states of excitations. The divergent damping of the excitations implied by the density of states also obviates a Bose-Einstein condensation. Were it to turn out that they are hard core bosons or neutral fermions, obvious modifications in the above hypotheses would be required. The possibility that the spectrum represents unfamiliar particles with unfamiliar statistics should also be entertained. The colossal degenerate fluctuating singlet state is likely to be unstable to other states by perturbations, for example superconductivity or itinerant charge states due to doping. 

The functional form of $\mathcal{A}_M(\omega,T)$ is derived at criticality in all the impurity models mentioned below in Sec.~\ref{sec:theory}, the 2D dissipative quantum $XY$ model \cite{Aji-V-qcf1, ZhuChenCMV2015, Hou-CMV-RG} and the SYK impurity model \cite{Sachdev93, Kitaev2017}. In every case, the low-energy excitations in these toy models are topological, as are the $T \to 0$ states.

 The free energy $F$ is given by
\begin{equation}
F = -kT \ln \mathrm{Tr}\,Z\,, \quad \mathrm{Tr}\,Z = \sum_{\lambda} e^{-\beta E_{\lambda}} \,, \
\end{equation}
where ${\lambda}$ contains the quantum numbers $\mathbf{q}$ as well as their occupation number. Summing over the occupation numbers gives, as usual,
\begin{equation}
 \ln \mathrm{Tr}\,Z = \sum_\mathbf{q} \ln \left(\frac{1}{2\sinh(\beta \omega_\mathbf{q}/2)}\right) \,.
\end{equation}

 Let us first calculate the entropy due to $\mathcal{A}_0(\omega,T)$. One finds the entropy from the free energy [$S = -(\partial F/\partial T)_V$]: \begin{align}
S & = k_B \frac{d}{dT} T \int d\omega \mathcal{A}_0(\omega,T) \ln \left(\frac{1}{2\sinh(\beta \omega/2)}\right), \\
\label{gsentr}
 & = k_B S_0 \int_0^\infty dx~x \frac{e^{-x}}{1+x^2} f(x) \,, \\
 &\quad f(x) = \ln \left[\textstyle{\frac{1}{2}} \csc(x/2)\right]\,.
\end{align}
The value of the integral is approximately 0.275.

The local density of states function $\mathcal{A}_M(\omega, T)$ gives a free energy proportional to $T^2$ with logarithmic corrections at high temperatures, an entropy of the deduced form, and a measurable specific heat
\begin{equation} \label{eq:cmovert}
C_M(T)/T \approx \gamma_M k_B 
\end{equation}
at low temperatures, with a logarithmic cutoff for temperatures above $T_x$. 

The two layers in LCSO complicate the analysis slightly. From the fits of Sec.~\ref{sec:2sublatt}, $C_M(T)/T$ values for the two layers differ by a factor of about 5. This is roughly the ratio of $\Theta_W$ values, which associates the lower $\Theta_W$ and hence the larger $C_M/T$ with the Cu1 layer. From the fit, for this layer $C_M(T)/T =$ 0.92~J~K$^{-2}$ per mole Cu, which yields $\gamma_M \approx 8.4\times10^{-13}$ s. 
 
The general form of the muon spin relaxation rate is
\begin{equation}
\label{lambda}
\lambda(T) \propto \gamma_{\mu}^2 \lim_{\omega \to 0} \frac{T}{\omega} \sum_\mathbf{q} | B_\mathrm{loc}(\mathbf{q})|^2 \mathrm{Im}\, \chi(\mathbf{q}, \omega) \,,
\end{equation} 
where $\sum_\mathbf{q} (T/\omega)\mathrm{Im}\,\chi(\mathbf{q},\omega)$ is the spectrum of magnetic fluctuations $\mathcal{A}_M(\omega,T)$ at the muon Zeeman frequency and is the correlation time $\tau_c$ in Eq.~(\ref{eq:lambda1}) \cite{Yaouanc11}; the same fluctuations are involved in both quantities. For $\mathcal{A}_M(\omega,T)$ given by Eq.~(\ref{eq:AM}), the temperature dependence of $\lambda(T)$ is seen to be the same as that of $C_M/T$, as in the experimental results shown in Fig.~\ref{fig:muSRspht}. Since $\mathcal{A}(\mathbf{q},\omega)$ is the absorptive part of the magnetic fluctuation spectra, it follows that at low temperatures $\tau_c \approx \gamma_M$. 

From the low-temperature muon spin relaxation rate, $\tau_c$ is estimated at ${\sim}1\text{--}2 \times 10^{-12}$ s (Sec.~\ref{sec:musr}). This is in fair agreement with the specific-heat $\gamma_M$ discussed above, but must be considered semi-quantitative at best for two reasons: $\gamma_M$ from $C_M(T)/T$ and $\tau_c$ from $\lambda(T)$ are only approximately equal, and the muon sites in the crystal structure are not known accurately.

\subsection{Survey of theories on spin-liquid models}
\label{sec:theory}

Detailed numerical calculations for the ground state of the $S{=}1/2$ nearest-neighbor AFM Heisenberg model (AHM) on a triangular lattice \cite{MeCi10} give an ordered three-sublattice ground state, with reduction of the order parameter by zero-point fluctuations of about 36\% for the nearest-neighbor interaction model. High-order high-temperature series expansions for the AHM \cite{ESY93, BeMi01} give a peak in the specific heat and the beginning of a decrease in the susceptibility at about 0.5 the exchange energy. As already mentioned, these bear no resemblance to our results. 

Numerical calculations on models with substantial 2nd-nearest-neighbor AHM interactions on a triangular lattice \cite{Hu19, CQLC19} have given a quantum-disordered state with gapless excitations conjectured to be spinons, but no ground-state entropy or indications of a gigantic peak in exponentially low-energy singlet excitations. Spinons have a Fermi surface and therefore a linear-in-$T$ specific heat, but the magnetic fluctuations associated with the Fermi surface lead to a relaxation rate proportional to the density of thermal excitations. It is therefore proportional to $T$ at low temperatures, similar to the Korringa rate in metals and unlike the constant rate found here. Models with ring-exchange \cite{SMZM20} are purported to give a chiral spin-liquid state but no low-energy singlet states. However, more work is needed on both these variations on the AHM.

The $S{=}1/2$ Heisenberg model on a kagom\'e lattice is considered the most likely spin liquid. Very careful calculations \cite{Misguich05} and analysis of a high-temperature series expansion up to 17th order for this model have found a missing entropy of about 1/2 the total value down to temperatures of $O(J/10)$, the lowest to which the calculations are reliable. These calculations have been further substantiated \cite{[{}] [{ and references therein.}] YUE20}. The results are similar to what we have found, but it is not known why the triangular and kagom\'e-lattice Heisenberg models should be in the same universality class.
 
It is worth noting that the Kitaev model in a certain range of parameters is similar to the toric model, which with an extreme choice of parameters exhibits a massive density of low-energy excitations followed by a continuum \cite{ChNu08, YNKM17}. The specific heat is proportional to $T^2$ below the peak in low-energy excitations from two dimensional spinons, but approximately linear in $T$ above that scale. Some gauge theory models in the large-$N$ limit, the so-called $\mathbb{Z}_2$ spin liquids \cite{ZKN17}, have similar properties.

Calculations on models for ice \cite{Pauling} or spin ice \cite{Anderson1956} (which agree with experiments \cite{GiSt36, Ramirez94}), and glass or spin-glass models, possess ground-state entropy but are obviously inapplicable here. Holographic field-theory models \cite{FILM10, Kachru2011, Zaanen} do have ground-state entropy as well as observable specific heat with various power laws including linear. (0+1)-dimensional disordered effective-impurity models such as the Sachdev-Ye-Kitaev (SYK) model \cite{Sachdev93, Kitaev2017} also have extensive ground-state entropy as well as gapless fermion excitations giving a linear-in-$T$ specific heat. 

The mapping of the SYK model to AdS theory of black holes has been discussed \cite{Kitaev2017}. Black holes are conjectured to be quantum-mechanical, and their physics is fashioned parallel to the thermodynamic laws \cite{BCH73}. They are believed to have an observable linear-in-$T$ entropy \cite{Carlip2015, Maldacena18u}. 
 
\subsection{Comparison with other spin liquids}
\label{sec:compare}

Notable experimental discoveries of compounds that do not order (or do not order down to very low temperatures compared to their $\Theta_W$'s) have been discussed as spin liquids for the last 25 years. We argue that our results (missing entropy, scale invariance) are unique to LCSO, most likely because it is purer than any other compound in its class of spin liquids so that its intrinsic properties are revealed. I.e., there may well be undiscovered missing entropy in other spin liquids that is obscured by lack of such purity.

In the following only a few representative references for each compound are given. More complete references may be found in reviews \cite{Savary16, BCKN20}.

1. Herbertsmithite is a $S{=}1/2$ kagom\'e lattice compound that has been thoroughly investigated \cite{dVKKS-B08, HNWR-R16, Norm16}. It has been found experimentally to have scale-invariant magnetic excitations, but the compound suffers from substantial disorder so that the pure limit specific heat has not been determined. It is reported to be $\propto T^{\alpha}$, $\alpha \approx 0.6$, at low temperatures, but this is obtained only after a contribution believed to be exclusively due to impurities is eliminated by applying a magnetic field of 9 T\@. No determination of magnetic entropy is available because a nonmagnetic counterpart, needed to subtract the lattice specific heat, has not been found. The compound cannot be made with less than a few percent Zn/Cu site disorder, leading to a Schottky specific heat. The low-temperature specific heat in zero-field is dominated by impurities and a field of 9 T reduces the specific heat to a power law $T^{\alpha}$, $\alpha \approx 0.6$. This is conjectured to represent the ``intrinsic'' specific heat, but its dependence on magnetic field cannot be determined.

The $\mu$SR rate is constant at low temperatures. It is interesting that single crystals have been made on which neutron scattering reveals a momentum independent continuum extending down to 0.25 meV, with $\omega/T$ scaling proposed here and much earlier in the physics of the cuprates. NMR and $\mu$SR results are similar to those in our compound, except at very low temperatures where evidence for inhomogeneity is found. 

2. Some organic Cu compounds, $\kappa$-(ET)$_2$Cu$_2\-$(CN)$_3$ \cite{YNOO08}, EtMe$_3$Sb[Pd(dmit)$_2$]$_2$ \cite{yamashita2011gaplessspin, NPSH19} have a linear-in-$T$ contribution to the heat capacity at low temperatures which, unlike our results, is field independent and is followed at higher temperatures by a large bump. We have not located any report in the literature of missing entropy or constant NMR rates at low temperatures. In earlier samples a linear-in-$T$ thermal conductivity was observed, but in more recent samples this is not found. The problems and different results in differently prepared samples have been documented in review articles, e.g. \cite{Savary16}.

3. Cs$_2$CuCl$_4$ \cite{CTTT01}, ZnCu$_3$(OH)$_6$Cl$_2$ \cite{HHCN12}, and BaCo$_2$(P$_{1-x}$V$_x$)$_2$O$_8$ \cite{ZCKN18} all show ordering of one or the other kind at low temperatures and have Curie contributions to the susceptibility shown by the authors to be due to several percent orphan spins.

4. YbMgGaO$_4$ \cite{LLZL15} exhibits a weakly divergent specific heat coefficient ($C/T \propto T^{-0.3}$) with less than 0.6\% residual spin entropy. But a low-temperature Curie tail is observed in the susceptibility, indicative of impurities that are not taken into account in the specific heat analysis. Mg-Ga site interchange is an intrinsic source of defects. The $\mu$SR relaxation rate is constant below $\sim$0.1~K, and does not track $C/T$ \cite{LABB16, DZZT20}.

5. Ba$_3$CuSb$_2$O$_9$ \cite{ZCLB11} has a nominally triangular $S{=}1/2$ lattice. Entropy saturation to only about 1/3 of $R \ln 2$ was observed using measurements on a non-magnetic analog to subtract the lattice contribution. But the sample has 5\% orphan spins. A huge peak in $C_M(T)/T$ occurs at about 6~K, and $C_M/T$ is field independent to 9 Tesla even between 0.2~K and 1~K well below $\mu_B H/k_B$. All this is quite different from the properties of nearly orphan-spin-free triangular lattice LCSO\@. Indications are that in Ba$_3$CuSb$_2$O$_9$ there is a collective state with glassy ordering at about 6~K, probably induced by non-local effects due to the large concentration of orphan spins.

6. NiGa$_2$S$_4$ \cite{TIKI08} and LCSO have very different properties. Spin freezing is observed in NMR experiments in the former below about 10~K\@. Relaxation-rate temperature dependencies have various power laws unrelated to the constant shown in LCSO.

7. TbInO$_3$ \cite{KWHW19} exhibits two distinct Tb ion sites due to a ferroelectric distortion. One forms a triangular lattice. No ordering is seen down to 0.15~K, but other properties do not resemble those in LCSO.

\section{CONCLUSIONS} \label{sec:concl}

In summary, two related phenomena have been discovered in the nearly defect-free compound LCSO. 

1. Conclusive evidence is found for missing entropy from a colossal density of singlet excitations below an ultra-low-energy scale compared to the Weiss temperature.  If these were local singlets, they would polarize and lose entropy in a magnetic field larger than the singlet-triplet separation. The fact that they are not removed by a magnetic field as high as 9~T shows that they are in a collective state where they cannot be addressed individually. At low temperatures there must be an insurmountable barrier against modification of this singlet characteristic. We can speculate that as in other systems, such as crystal dislocations or superconducting vortices, the barrier is topological in nature. 

2. Quantitatively related magnetic specific heats $C_M/T$ and $\mu$SR relaxation rates $\lambda(T)$ are observed below a temperature related to the Weiss temperature $\Theta_W$, followed by the same logarithmic cutoff in both measurements at higher temperatures. The excitations necessary for these are shown to be scale invariant. They carry finite spin quantum numbers because their entropy for $g\mu_B H \lesssim k_B T$ is systematically reduced in a field; this leads to temperature-independent muon spin relaxation. There are no other {\it measurable} excitations at any temperature and up to 9 tesla. All measured properties can be related to a single parameter in the scaling function. 

Both the specific heat and the $\mu$SR relaxation rate follow from the scale-invariant density of states function $\mathcal{A}_M(\omega, T) = \gamma_M f(\omega/T)$ for magnetic fluctuations described in Sec.~\ref{sec:invar}. Not only is the temperature dependence of both quantities given by this form, but their orders of magnitude are obtained from the same coefficient $\gamma_M$. As a function of imaginary time $\tau$ periodic in inverse temperature, $\mathcal{A}_M(\omega,T)$ is equivalent to an algebraic decay $\propto 1/\tau$. A ground-state entropy, which should more accurately be called a temperature-independent entropy, requires a more singular form $\mathcal{A}_{0} (\omega, T)$, which corresponds to a correlation function of the singlets approximately proportional to $1/\log(\tau)$. This is as quantum as one can get. Some conceptual questions related to this are briefly discussed in SI Sec.~IX\@. This form is chosen in the belief that the missing entropy is due to a dynamical effect. The form can be modified easily by introducing a new scale if instead there are equally unexpected colossal ultra-low-energy excitations.

Theoretical results for spin liquids and their relation to our experimental findings are briefly summarized in Sec.~\ref{sec:theory}. We have not found theoretical results on any relevant model which correspond to the properties discovered here \cite{Savary16, BCKN20}. Similarly, in a close look at the literature (a summary is given in Sec.~\ref{sec:compare}), we find that such properties have not been previously observed in any spin-liquid candidates. We think this is because LCSO can be prepared with fewer defects than any other spin liquid investigated so far, so that the intrinsic behavior of its class of spin liquids is revealed. The simplicity and the nature of the singularities in Eqs.~(\ref{eq:Aloc})--(\ref{eq:AM}), with which we can parameterize all the data, invite important new theoretical developments. The magnetic fluctuations suggested by $\mathcal{A}_M(\omega,T)$ should be accessible via neutron scattering. The detection of the scalar excitations $\mathcal{A}_0(\omega,T)$ poses an interesting challenge to experimental techniques. Having no charge or magnetic moment, they are a form of dark matter not observable by the usual spectroscopic techniques.

\vspace{20pt}\section*{Acknowledgments}

We are grateful to Patrick A. Lee for extensive discussions of this work and for suggesting temperature scaling to estimate the high-temperature magnetic lattice specific heat, and to Rajiv R. P. Singh for providing his results for the high-temperature series expansions. We thank B. Hitti and D. J. Arseneau of TRIUMF and the staff of the Paul Scherrer Institute for their valuable help during the $\mu$SR experiments. C.M.V. performed this work while a ``Recalled Professor'' at UC Berkeley, and wishes to thank the members of the condensed-matter theory group for their hospitality. This research was funded by the National Natural Science Foundations of China, No.~12034004 and No.~11774061, and the Shanghai Municipal Science and Technology (Major Project Grant No.~2019SHZDZX01 and No.~20ZR1405300). 

\vspace{15pt} \bibliography{LCSOv7-2.bib}

\end{document}